\documentclass[nofootinbib, onecolumn, longbibliography, superscriptaddress,amsmath,amssymb,amsfonts]{revtex4-2} 

\usepackage{graphicx}
\usepackage{color}
\usepackage{qcircuit}
\usepackage{braket}
\usepackage{bbm}
\usepackage{bm}
\usepackage{comment}
\usepackage{hyperref}
\usepackage{cleveref}
\usepackage{lipsum}
\usepackage[frozencache,cachedir=minted-cache]{minted}
\usepackage{xcolor}
\usepackage{mdframed}
\definecolor{LightGray}{gray}{0.95}

\crefname{equation}{Eq.}{Eqs.}

\newcommand{\beq}{\begin{equation}}
\newcommand{\eeq}{\end{equation}}

\begin{document}

\title{Grad DFT: a software library for machine learning enhanced density functional theory}

\author{Pablo A. M. Casares}
\email{pablo.casares@xanadu.ai}
\affiliation{Xanadu, Toronto, ON, M5G2C8, Canada}
\author{Jack S. Baker}
\affiliation{Xanadu, Toronto, ON, M5G2C8, Canada}
\author{Matija Medvidovi\'c}
\affiliation{Xanadu, Toronto, ON, M5G2C8, Canada}
\affiliation{Center for Computational Quantum Physics, Flatiron Institute, New York, NY, 10010, USA}
\affiliation{Department of Physics, Columbia University, New York, 10027, USA}
\author{Roberto dos Reis}
\affiliation{Xanadu, Toronto, ON, M5G2C8, Canada}
\author{Juan Miguel Arrazola}
\affiliation{Xanadu, Toronto, ON, M5G2C8, Canada}
\date{\today}

\begin{abstract}

Density functional theory (DFT) stands as a cornerstone method in computational quantum chemistry and materials science due to its remarkable versatility and scalability. Yet, it suffers from limitations in accuracy, particularly when dealing with strongly correlated systems. To address these shortcomings, recent work has begun to explore how machine learning can expand the capabilities of DFT; an endeavor with many open questions and technical challenges. In this work, we present Grad DFT: a fully differentiable JAX-based DFT library, enabling quick prototyping and experimentation with machine learning-enhanced exchange-correlation energy functionals. Grad DFT employs a pioneering parametrization of exchange-correlation functionals constructed using a weighted sum of energy densities, where the weights are determined using neural networks. Moreover, Grad DFT encompasses a comprehensive suite of auxiliary functions, notably featuring a just-in-time compilable and fully differentiable self-consistent iterative procedure. To support training and benchmarking efforts, we additionally compile a curated dataset of experimental dissociation energies of dimers, half of which contain transition metal atoms characterized by strong electronic correlations. The software library is tested against experimental results to study the generalization capabilities of a neural functional across potential energy surfaces and atomic species, as well as the effect of training data noise on the resulting model accuracy.

\end{abstract}

\maketitle
\section{Introduction}

Density functional theory (DFT) is one of the most successful methods for simulating molecules and materials, largely due to its relatively low computational cost and ability to provide useful calculations across many areas of interest~\cite{hohenberg1964inhomogeneous,kohanoff2006electronic}. But DFT is also known for its limitations, particularly when dealing with systems whose behavior is dominated by strong electronic correlations~\cite{sie_dft, cohen2008insights, cohen2012challenges}. Throughout decades of work, many functionals have been proposed based on a combination of theoretical insights and fits to experimental data~\cite{mardirossian2017thirty}. Yet, it remains an outstanding challenge to construct functionals that are both inexpensive and highly accurate across a broad spectrum of systems~\cite{jain2016computational,peverati2014quest, mardirossian2017thirty, medvedev2017density}. \\

The remarkable success and rapid progress of machine learning techniques have motivated its use to advance DFT~\cite{snyder2012finding, brockherde2017bypassing, nagai2020completing, dm21, nagai2022machine, pederson2022machine}. The central premise involves leveraging the capabilities of deep learning models to systematically harness large amounts of data and train models with strong generalization potential~\cite{bartlett2021deep, belkin2021fit}. While there exists evidence from quantum computational complexity that an exact functional cannot be found for general problems without incurring exponential costs~\cite{schuch2009computational}, the availability of data provides a potential workaround in practice, since algorithms can be more powerful in the presence of data~\cite{mcclean2021foundations,huang2021power}. Therefore, it is conceivable that a combination of machine learning methods and high-quality data will pave the way for the next frontier of accuracy in DFT. Motivated by these insights, a few proof-of-principle designs for new machine learning based functionals have been proposed~\cite{li2016understanding,nagai2020completing,dick2020machine,ma2022evolving,chen2020deepks,kasim2021learning,tamayo2018automatic,laestadius2019kohn}.\\

\begin{figure}
    \centering
\includegraphics[width=0.7\columnwidth]{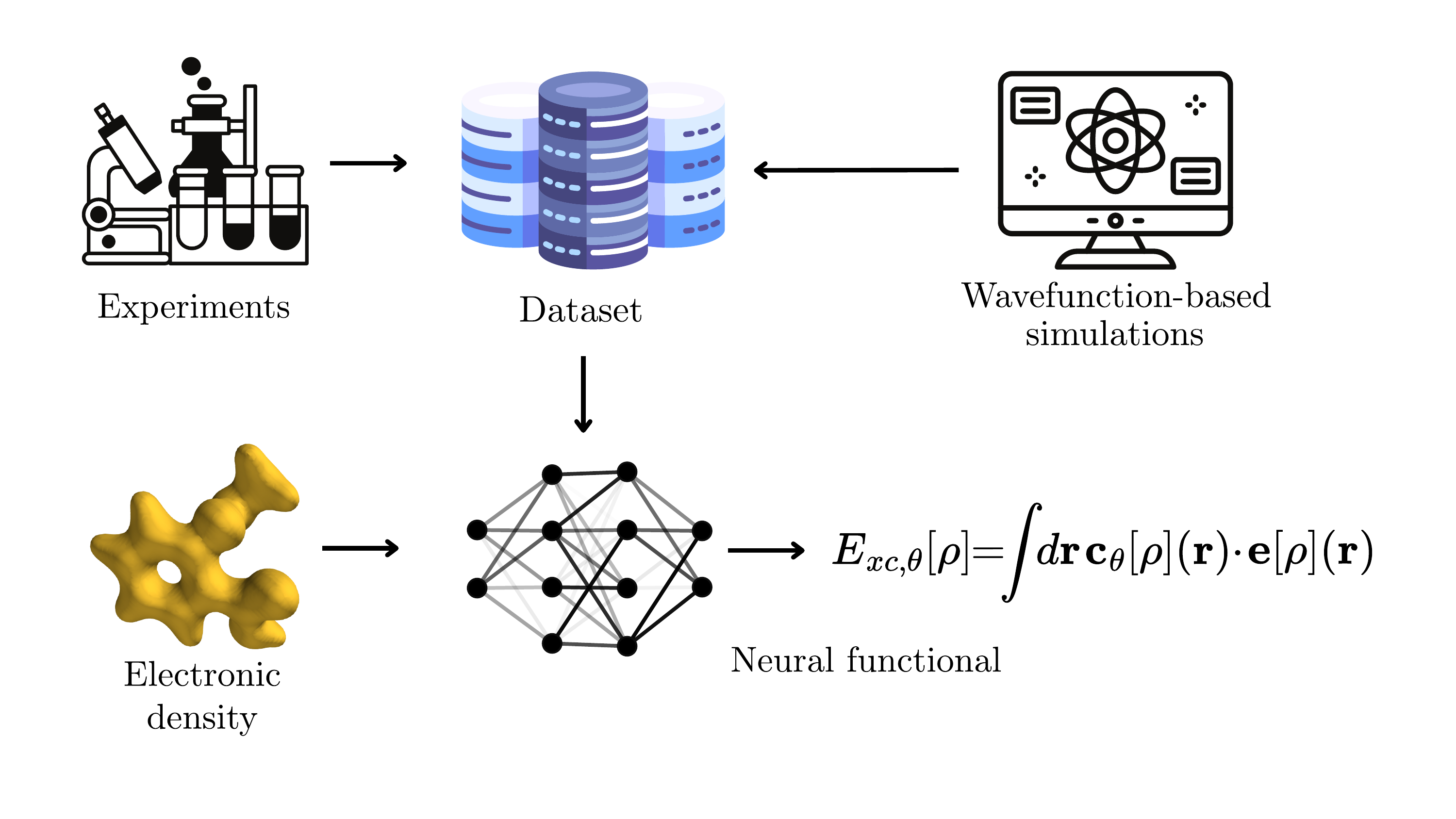}
    \caption{Schematic depiction of the general workflow for machine learning-enhanced density functional theory. We envision a setting where high-quality data is generated using experiments and advanced wavefunction-based simulations on classical or quantum computers. These act as data factories generating large datasets that can be used for training new functionals. We consider the case of \textit{neural functionals}, functionals in which neural networks are used to predict the local weights $\bm{c}_{\theta}[\rho](\bm{r})$ of associated energy densities $\bm{e}_{\theta}[\rho](\bm{r})$, when given the electronic density as input. $E_{xc,\theta}$ stands for a $\theta$-parametrized exchange-correlation energy functional.}
    \label{fig:introductory_figure}
\end{figure}

Success in combining machine learning and DFT will require many developments. We envision a general workflow, illustrated in Fig.~\ref{fig:introductory_figure}, where experiments, high-performance supercomputers --- and eventually quantum computers --- are used as data factories to generate large collections of accurate ground-state properties for challenging molecules and materials. These data are then leveraged by advanced software and algorithms to train exchange-correlation functionals using machine learning techniques. The desired outputs of this process are functionals whose accuracy can rival the methods used for data generation, while retaining the lower computational cost of DFT methods.  \\

In this work, we implement the software layer of this general workflow, introducing Grad DFT: a software library for designing and training differentiable functionals using machine learning techniques. Written in JAX~\cite{jax2018github}, the library is fully differentiable and deployable to high-performance computing (HPC) systems using multiple GPUs and TPUs. The code is open-source and publicly available at \url{https://github.com/XanaduAI/GradDFT}. We focus on functionals based on neural networks, which we refer to as \emph{neural functionals}. Grad DFT facilitates the implementation of a wide variety of differentiable functionals supporting LDA, GGA and meta-GGA approximations, hybrid schemes accounting for the exact exchange energy, and dispersion corrections. All this is complemented with auxiliary functions, such as fully differentiable self-consistent field (SCF) procedures including, among others, direct inversion in the iterative subspace, all compatible with just-in-time (JIT) compilation.
We also provide a diverse selection of loss function regularizers, encompassing fifteen of the known constraints of the universal functional~\cite{kaplan2023constraints}. Finally, we give examples of use as well as integration tests replicating well-known standard functionals and neural functionals including B3LYP~\cite{b3lyp} and DM21~\cite{dm21}. \\

The purpose of this manuscript is twofold. It is intended as a companion to the software library and presents state-of-the-art experimentation with machine learning-enhanced DFT leveraging the library. We first describe the mathematical strategy for parametrizing functionals, establishing the foundation for user-designed neural functionals. We then translate these concepts into the code architecture, guiding the user through example code snippets depicting its main features. In addition to the software library, we also compile a dataset of experimental dissociation energies of dimers indicated in Fig.~\ref{fig:dataset}. We employ this dataset in some of our numerical experiments studying the generalization capabilities of neural functionals and the impact of training data accuracy on the resulting model precision. \\

\section{Related work\label{sec:RelatedWork}}

Over the past decade, there have been many proposals on how DFT may benefit from machine learning techniques~\cite{li2016understanding,nagai2020completing,dick2020machine,ma2022evolving,chen2020deepks,kasim2021learning,tamayo2018automatic,laestadius2019kohn}. For instance, there exist differentiable versions of PySCF and LibXC~\cite{DiffPySCF, marques2012libxc,lehtola2018libxc,zheng2023jax}. Early research in the literature applied these techniques to one-dimensional systems~\cite{snyder2012finding,hollingsworth2018can,denner2020efficient,custodio2019artificial,schmidt2019machine,li2021kohn}, simple molecules~\cite{dick2020machine,kasim2021learning,tamayo2018automatic}, and to addressing the fractional electron problem~\cite{dm21}. Prior work has also explored a diverse range of approaches such as modeling the universal or kinetic energy functionals~\cite{ryczko2019deep,meyer2020machine,seino2018semi,seino2019semi,fujinami2020orbital,golub2019kinetic,alghadeer2021highly,ghasemi2021artificial,ellis2021accelerating}, adding corrections to existing exchange-correlation functionals~\cite{bogojeski2020quantum}, and modeling the exchange-correlation potential $v_{xc}[\rho](r) = \frac{\delta E_{xc}[\rho](r)}{\delta \rho(r)}$~\cite{nagai2018neural,ryabov2022application,schmidt2019machine}.
While most architectures are based on feed-forward neural networks, some other proposals have suggested using convolutional filters to go beyond the locality assumption~\cite{lei2019design,zhou2019toward}.\\

Other approaches have focused on the self-consistent loop; either to bypass it with a direct optimization of the orbitals~\cite{brockherde2017bypassing,chandrasekaran2019solving,li2023dft}, or use it during training to regularize the learned functional~\cite{li2021kohn,kalita2022well}. Finally, some effort has also been devoted to understanding how to incorporate known constraints of the exact exchange-correlation energy functional to its design~\cite{dick2021highly,bystrom2022cider, nagai2022machine, dm21}.\\

In contrast to the previous studies, our work does not aim to present a new functional. Rather, we introduce a software library designed to facilitate their creation using machine learning methods. In other words, our goal is to fill a gap in the field by providing a versatile tool for research and exploration.

\section{Parametrized functionals\label{sec:parametrized_functionals}}

In this section, we summarize strategies for constructing exchange-correlation functionals, and describe how they can be expressed in terms of trainable parameters. We review methods to include inhomogeneity factors, exact exchange contributions, and dispersion corrections. We then describe how the components of the functional can be combined using coefficient functions, typically obtained as outputs of a neural network. The result is a general template to build parametrized models for exchange-correlation functionals.\\

We focus on Kohn-Sham (KS) DFT, where the non-spin-polarized total energy functional $E_{\text{KS}}[\rho]$ is given by~\cite{kohn1965self}:
\begin{equation}
\begin{split}\label{eq:E_KS}
E_{\text{KS}}[\rho] = \sum_{i=1}^{\text{occ}}&\int d\bm{r}\; |\nabla \varphi_{i}(\bm{r})|^2 +\int d\bm{r}\; U(\bm{r}) \rho(\bm{r})  + \frac{1}{2}\int d\bm{r} d\bm{r}'\frac{\rho(\bm{r})\rho(\bm{r}')}{|\bm{r}-\bm{r}'|} + E_{II} + E_{xc}[\rho],
\end{split}
\end{equation}
where $\rho(r)$ is the ground-state electronic density, self-consistently determined by the (lowest-energy) occupied orbitals $\varphi_i(r)$. 
The first and second terms in Eq.~\eqref{eq:E_KS} are the KS kinetic and potential energy functionals, respectively, with $U(\bm{r})$ denoting the one-electron external potential, often the Coulomb potential generated by a collection of ions, nuclei if we do not use pseudopotentials. The third term is the classical electrostatic Hartree energy, and $E_{II}$ represents the Coulomb interaction between ions. Finally, $E_{xc}[\rho]$ is the contribution of the exchange and electron correlation effects to the total energy. It is this last term for which an exact expression is not known so approximations must be made. $E_{xc}[\rho]$ can be further separated into individual exchange and correlation components~\cite{martin2020electronic}:

\begin{align}\label{eq:separating_exchange_correlation}
    E_{xc}[\rho]  = \int d\bm{r} \left(e_{x}[\rho](\bm{r}) + e_{c}[\rho](\bm{r})\right).
\end{align}
Throughout this work, we use the term \emph{energy density} to refer to a function $e[\rho](\bm{r})$ that is integrated over space to calculate the energy, i.e., 
\begin{equation}
    E[\rho] = \int d\bm{r}\; e[\rho](\bm{r}).
\end{equation}
In Eq.~\eqref{eq:separating_exchange_correlation}, $e_{x}[\rho](\bm{r})$ is the exchange energy density and $e_{c}[\rho](\bm{r})$ is the correlation energy density. We adopt a convention where all explicit dependency on the electronic density $\rho(\bm{r})$ is absorbed inside the energy densities $e[\rho](\bm{r})$.\\

There are multiple ways to choose these energy densities. The simplest is the local (spin) density approximation (LDA), whose exchange energy density is given by
\begin{equation}
    \epsilon_{x,\sigma}[\rho](\bm{r}) = -\frac{3}{2}\left(\frac{3}{4\pi}
    \right)^{1/3}\rho_\sigma(\bm{r})^{4/3}.\label{eq:LDA_density}
\end{equation}
$\sigma$ denotes the spin projection quantum number, which can take values up ($\uparrow$) or down ($\downarrow$). The correlation energy density for LDA is not known in analytical form but it can be expressed in terms of an accurate fit~\cite{CeperleyAlder,vwn, pw92}. \\

To improve the quality of the functional, multiplicative inhomogeneity correction factors $F_x$ and $F_c$ can be introduced. These generally depend on the density and its higher-order derivatives. In this case, the exchange-correlation functional is given by

\begin{equation}\label{eq:xcfunctional}
\begin{split}
E_{xc}[\rho] &= E_x[\rho] +E_c[\rho] := \int d\bm{r} \left(e^F_x[\rho](\bm{r})+ e^F_c[\rho](\bm{r})\right),
\end{split}
\end{equation}
where we have implicitly defined inhomogeneous energy densities
\begin{align}
    e^F_x[\rho](\bm{r}) = F_{x}[\rho](\bm{r})  e_{x}[\rho](\bm{r}),\\
    e^F_c[\rho](\bm{r}) = F_{c}[\rho](\bm{r})  e_{c}[\rho](\bm{r}).
\end{align}
The inhomogeneity factors $F$ can be expanded as polynomials of dimensionless variables that depend on the gradients of the density~\cite[Eq.~13]{mardirossian2017thirty}:
\begin{equation}\label{eq:ICF}
    F_{\sigma}[\rho] = \sum_{n = 0}^N\sum_{m = 0}^M s_{nm} u_\sigma^n w_\sigma^m.
\end{equation}
The variables $u_\sigma$ and $w_\sigma$ must be bounded and are used in the Generalized Gradient Approximation (GGA) and meta-GGA functionals. They are expressed in terms of the dimensionless variables $x_\sigma$ and $t_\sigma$
\begin{align}
   x_\sigma &= \frac{|\nabla \rho_\sigma|}{\rho^{4/3}_\sigma},\label{eq:x_sigma}\\
   t_\sigma &= \frac{3}{5}(6\pi^2)^{\frac{2}{3}}\frac{\rho_\sigma^{5/3}}{\tau_\sigma}, \label{eq:t_sigma}
\end{align}
where the term
\begin{equation}\label{eq:tau_sigma}
    \tau_\sigma = \frac{1}{2}\sum_{i=1}^{\text{occ}} |\nabla \varphi_{\sigma,i}|^2,
\end{equation}
represents the kinetic energy of the KS system. We removed the explicit dependency on $\bm{r}$ in~\cref{eq:x_sigma,eq:t_sigma,eq:tau_sigma} for notational clarity. The variables $u_\sigma$ and $w_\sigma$ can be constructed in different ways to ensure their bounded range, for example as
\begin{align}
    u_\sigma &= \frac{\beta x_\sigma^2}{1+ \beta x_\sigma^2},\\
    w_\sigma &= \frac{t_\sigma-1}{t_\sigma+1}.
\end{align}

Inhomogeneous corrections can be generalized to include higher-order derivatives, although this is less common. In this case, the factors would be expressed as
\begin{equation}\label{eq:ICF_polynomial_generalization}
    F_{\sigma}[\rho]=\sum_{n_1\ldots n_k} s_{n_1\ldots n_k;\sigma} u_{1,\sigma}^{n_1}\ldots u_{k,\sigma}^{n_k}.
\end{equation}
The variables $u_j$ can be any finite-range dimensionless variable of the $j$-th derivative of the electronic density. One possible example, omitting the spin variable, is~\cite{martin2020electronic}
\begin{equation}\label{eq:u_j}
    u_j = \frac{\beta x_j}{1+\beta x_j};\qquad x_j = \frac{|\nabla^j \rho|}{2^j(3\pi^2)^{j/3}\rho^{1+j/3}},
\end{equation}
with $\beta>0$. Expressions to combine the spin polarizations are known for both the exchange and correlation energy densities~\cite{ullrich2011time}.\\

The functional in Eq.~\eqref{eq:xcfunctional} is parametrized in terms of the coefficients $s_{nm}$ entering Eq.~\eqref{eq:ICF}, which can be chosen to be functions of the spatial coordinate, $s_{nm} = s_{nm,\theta}(\bm{r})$. Furthermore, they may also depend on additional variables $\theta$, which become trainable parameters of the model. The resulting functional can be expressed as

\begin{equation}
    E_{xc, \theta}[\rho] = \int d\bm{r} \left(e^F_{x, \theta}[\rho](\bm{r})+ e^F_{c, \theta}[\rho](\bm{r})\right). 
\end{equation}

The model can also treat hybrid functionals, which utilize the exact expression for the exchange interaction
\begin{equation}\label{eq:HF_component}
\begin{split}
    &E^{\text{HF}}_{\omega,\sigma} =\int d\bm{r} \; e^{\omega\text{HF}}_{\sigma}[\rho](\bm{r}),
\end{split}
\end{equation}
where the range-separated Hartree-Fock energy density is given by
\begin{align}\label{eq:range_separated_HF}
    &e^{\omega\text{HF}}_{\sigma}[\rho](\bm{r}) = \sum_{pqrs}\frac{\Gamma^\sigma_{pq}\Gamma^\sigma_{rs}}{2}\int d\bm{r}'\phi_p(\bm{r})\phi_s^*(\bm{r})\frac{\text{erf} (\omega|\bm{r}-\bm{r}'|)}{|\bm{r}-\bm{r}'|}\phi_r(\bm{r}')\phi_q^*(\bm{r}').
\end{align}
Here $\Gamma^\sigma_{pq}$ is the one-particle reduced density matrix that can be used to express the electronic density in terms of the atomic orbitals $\phi(\bm{r})$ as follows,
\begin{equation}
    \rho_{\sigma}(\bm{r}) = \sum_{pq}\Gamma^\sigma_{pq} \phi_{p}(\bm{r})\phi_{q}^*(\bm{r}).
\end{equation}
The parameter $\omega > 0$ regularizes the Coulomb singularity in Eq.~\eqref{eq:range_separated_HF} when $\bm{r} = \bm{r}'$. The Coulomb kernel is recovered in the limit $\omega \rightarrow \infty$, a case we denote by $e^{\text{HF}}$. It is also possible to include other wavefunction components, based on, for example, perturbation theory~\cite{grimme2006mp2dft}, the random phase approximation, or even coupled-cluster~\cite{kalai2019range}, in which case they are called double-hybrid functionals~\cite{goerigk2014double}.\\

\begin{figure}
    \centering
    \includegraphics[width=\textwidth]{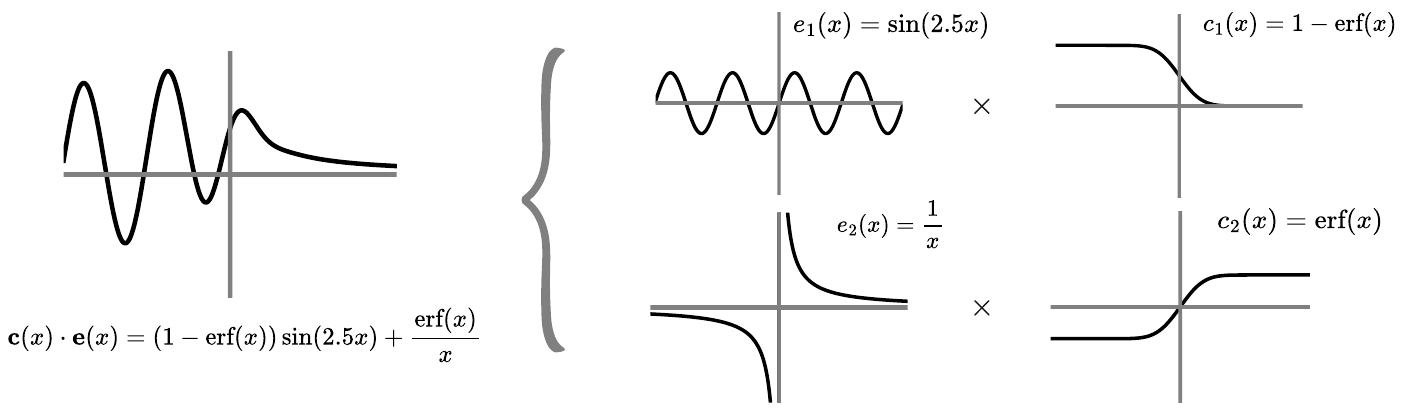}
    \caption{Conceptual representation of the importance of allowing the coefficients $\bm{c}(\bm{r})$ to vary over space. In this example, we consider a basis of two one-dimensional functions $e_1(x)=\sin(2.5x)$ and $e_2(x)=1/x$. The goal is to combine them to produce the target function on the left, which transitions from sinusoidal oscillations to monotonic decay. Replicating this behavior is not possible choosing constant coefficients $c_1$ and $c_2$, but it can be achieved by selecting associated coefficient functions $c_1(x)=1-\text{erf}(x)$ and $c_2(x)=\text{erf}(x)$, and setting the model to be $\bm{c}(x)\cdot \bm{e}(x)=c_1(x)e_1(x) + c_2(x)e_2(x)$.}
    \label{fig:coefficients}
\end{figure}

Finally, the functional may include corrections due to dispersion effects such as DFT-D tails~\cite{grimme2004accurate,grimme2006semiempirical,grimme2010consistent,grimme2011effect}. These are given by a direct contribution to the energy of the form
\begin{equation}\label{eq:dispersion}
    E_{\text{DFT-D}} = -\sum_{I<J}\sum_{n = 3, 4,\ldots} \frac{\alpha^{D, IJ}_{2n}}{R_{IJ}^{2n}}f_{2n}(R_{IJ}).
\end{equation}
$R_{IJ} = |\bm{R}_{I}-\bm{R}_{J}|$ indicates the distance between nuclei $I$ and $J$, the coefficients $\alpha^{D,IJ}_{2n}$ are parameters, and $f_{2n}$ represent damping functions. For example, in Ref. ~\cite{grimme2011effect}
\begin{equation}\label{eq:damping_dispersion}
    f_{2n}(R_{IJ}) = \frac{1}{1 + e^{-\gamma(R_{IJ}/R_{IJ}^0-1)}},
\end{equation}
with $R_{IJ}^0$ a scaling factor and $\gamma$ a steepness parameter. An alternative is the use of non-local correlation functionals such as VV10~\cite{vv10},
\begin{align}\label{eq:VV10}
    E_{\text{VV10}}[\rho] &=  \int d\bm{r}'d\bm{r} \;\rho(\bm{r}) \Phi(\bm{r},\bm{r}')\rho(\bm{r}'):= \int d\bm{r} \;e^{\text{VV10}}[\rho](\bm{r}),
\end{align}
with $\Phi$ denoting a two-body parametrized potential dependent on $\frac{|\nabla \rho|}{\rho}$ at $\bm{r}$ and $\bm{r}'$, and their separation $r = |\bm{r}-\bm{r}'|$.\\

\begin{figure}
    \centering
    \includegraphics[width=0.8\textwidth]{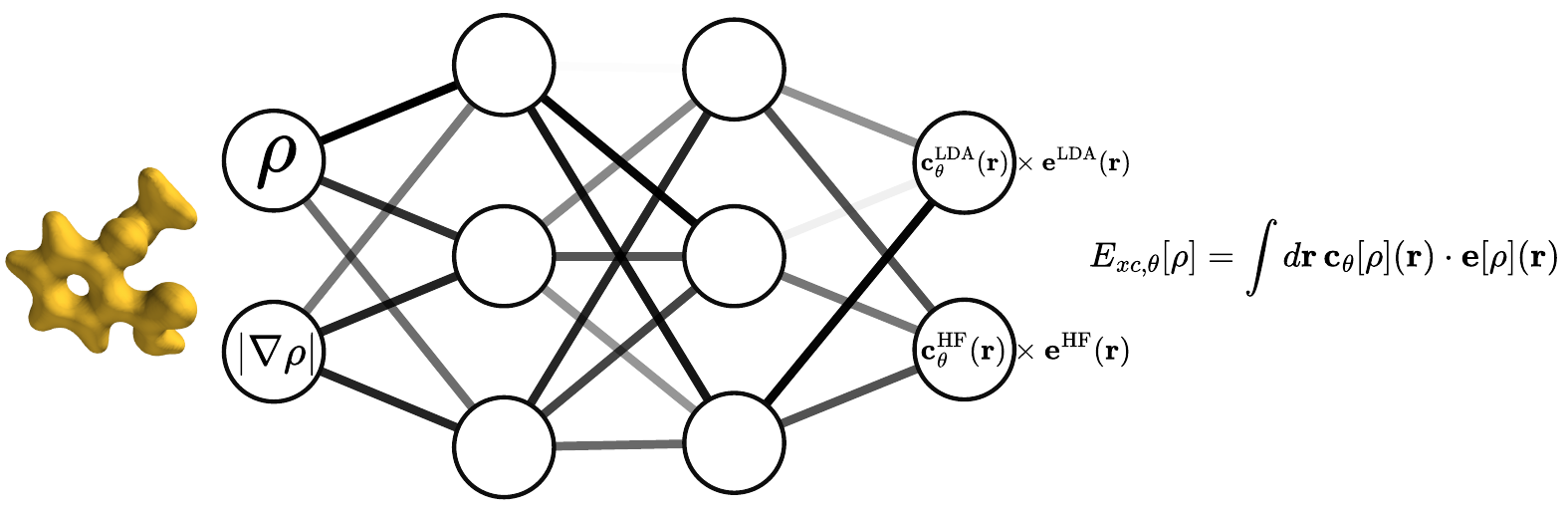}
    \caption{A pictorial summary and example of how the neural functional works. The value of the coefficients $\bm{c}_\theta[\rho](\bm{r})$ is fixed by a neural network whose input is the electronic density, its gradients, etc. The coefficients get dot-multiplied by some energy densities, and the result is integrated to compute the exchange-correlation energy.}
    \label{fig:neural_functional}
\end{figure}

In summary, functionals are usually constructed as a combination of different individual functionals, typically approximate exchange and correlation components, exact exchange contributions, and dispersion corrections~\cite{mardirossian2017thirty}. Each such functional $E_i[\rho]$ may have free internal parameters, and can be expressed in terms of a corresponding energy density $e_i[\rho](\bm{r})$. It is helpful to view the functionals $E_{i}[\rho]$, and their associated energy densities $e_i[\rho](\bm{r})$ as a basis. This means we can choose coefficients $c_i$ to expand the exchange-correlation functional as
\begin{equation}
    E_{xc}[\rho] = \int d\bm{r}\; \sum_{i=1}^K c_i\, e_{i}[\rho](\bm{r}) =  \sum_{i=1}^K c_i E_{i}[\rho].
\end{equation}
More generally, coefficients can be parametrized functions of space that depend also on the density and its derivatives, i.e., $c_i = c_{i,\theta}[\rho](\bm{r})$. We refer to these as \emph{coefficient functions}, and their use significantly expands the flexibility of the functional parametrization; see Fig.~\ref{fig:coefficients} for a conceptual example. We employ compact notation and define the vectors
\begin{align}
    \bm{c}_{\theta}[\rho](\bm{r})&=\left(c_{1,\theta}[\rho](\bm{r}), c_{2,\theta}[\rho](\bm{r}), \ldots, c_{K,\theta}[\rho](\bm{r})\right);\label{eq:vec_coeffs}\\
    \bm{e}[\rho](\bm{r})&=\left(e_1[\rho](\bm{r}), e_2[\rho](\bm{r}), \ldots, e_K[\rho](\bm{r})\right),\label{eq:vec_densities}
\end{align}
which capture all coefficient functions and energy densities. The exchange-correlation functional can then be generally written as 
\begin{equation}\label{eq:param-func}
    E_{xc, \theta}[\rho] = \int d\bm{r} \,\bm{c}_{\theta}[\rho](\bm{r}) \cdot \bm{e}[\rho](\bm{r}).
\end{equation}
The Grad DFT software library presented in this work is based on the general expression of Eq.~\eqref{eq:param-func}. This is illustrated in Fig.~\ref{fig:neural_functional}. We focus on energy densities and coefficient functions as the central abstractions to design functionals, and place an emphasis on neural functionals, for which the vector $\bm{c}_\theta[\rho](\bm{r})$ is obtained as the output of a neural network. These generally take the density and other derived quantities as input.\\

Most density functionals are specific cases of this general model. For instance, the functional $\omega$B97M-V~\cite{wB97M-V} is based on expression~\eqref{eq:ICF} with $N = 4$
and $M = 8$, range-separated Hartree-Fock, and a VV10 non-local correlation dispersion. The coefficients and weights in Eq.~\eqref{eq:ICF} and the values of $N$ and $M$ were combinatorially optimized to fit experimental data. Another example is DM21~\cite{dm21}. Their coefficient functions are produced as the output of a neural network:
\begin{equation}\label{eq:DM21}
    E_{xc}^{\text{DM21}}[\rho] = \int  \bm{c}_{\theta}^{\text{DM21}}[\rho](\bm{r})\cdot
    \begin{bmatrix}
    e_x^{\text{LDA}}[\rho](\bm{r})\\
    e^{\text{HF}}[\rho](\bm{r})\\
    e^{\omega \text{HF}}[\rho](\bm{r})\\
    \end{bmatrix}
    d\bm{r}.
\end{equation}

Our library specializes in allowing users to design differentiable functionals of the form in Eq.~\eqref{eq:param-func}, and to train their parameters using gradient-based optimization. In the following section, we provide more details on the capabilities of the library, with example code.

\section{The Grad DFT library }\label{sec:ML4DFT}


Our software is a JAX-based library designed for rapid prototyping and experimentation with differentiable functionals, with a specific focus on neural functionals~\cite{jax2018github}. It is designed to operate with libraries in the JAX ecosystem such as Flax for building neural networks~\cite{flax2020github}, Optax for managing optimizers and training schedules, and Orbax for saving and loading checkpoints~\cite{JaxEcosystem}. The checkpoints record the value of the parameters of the neural network during training, as well as the state of the optimizer. On the chemistry side, seamless integration with  PySCF~\cite{sun2018pyscf,sun2020recent} greatly streamlines the data preprocessing. The primary workflow is outlined in the box below.\\

\begin{mdframed}[nobreak=true]
    Workflow of the library.
    \begin{enumerate}
        \item Specify an instance of the \texttt{Molecule} class, which has methods to compute the electronic density $\rho$ and derived quantities.
        \item Define the function \texttt{energy\_densities}, to calculate $\bm{e}[\rho](\bm{r})$.
        \item Implement the function \texttt{coefficients}, which may include a neural network, to compute $\bm{c}_\theta[\rho](\bm{r})$.
        \item Create an instance of the \texttt{Functional} class, and use
        \texttt{functional.energy(molecule, params)} to compute the total energy $E_{\text{KS}}[\rho]$ non self-consistently. The argument \texttt{params} indicates neural network parameters $\theta$ used to evaluate the exchange-correlation energy functional,
        $$E_{xc,\theta}[\rho] = \int d\bm{r} \bm{c}_\theta[\rho](\bm{r})\cdot\bm{e}[\rho](\bm{r}).$$
        \item Train the neural functional using JAX automatic differentiation capabilities, in particular~\texttt{jax.grad}.
    \end{enumerate}
\end{mdframed}
\vspace{0.4cm}

The code is built on top of two main classes: \texttt{Molecule} and \texttt{Functional}. Objects of the \texttt{Molecule} class store the properties of chemical systems as attributes, that can then be used in differentiable calculations. These include the reduced density matrix, the electronic repulsion tensor, and the molecular orbital coefficients, among others. The user may compute all these arrays manually, or make use of the function \texttt{molecule\_from\_pyscf} to instantiate a \texttt{Molecule} object directly from PySCF. The code below demonstrates how to build a molecule object:\\
\begin{minted}[bgcolor=LightGray, linenos]{python}
from grad_dft import molecule_from_pyscf
from pyscf import gto, dft
# Define a PySCF mol object for the H2 molecule
mol = gto.M(atom=[['H', (0, 0, 0)], ['H', (0.74, 0, 0)]], basis='def2-tzvp', spin=0)
# Create a PySCF mean-field object
mf = dft.UKS(mol)
mf.kernel()
# Create a Molecule from the mean-field object
molecule = molecule_from_pyscf(mf)
\end{minted}
Using this function allows users to benefit from all options available in PySCF for constructing molecule objects, such as the choice of basis set and spin multiplicity. \\ 

Of importance among the attributes of the molecule object is the associated grid, which is generated to discretize the space over which the electronic and energy densities are evaluated. 
The class \texttt{Molecule} provides methods to compute these quantities. For example, \texttt{molecule.density()} outputs the electronic density and \texttt{molecule.kinetic\_density()} returns the kinetic energy density.
Additionally, we provide functionalities to save and load these objects in HDF5 format~\cite{hdf5}. This is key given the memory footprint of some of the arrays managed by the code.\\

Instantiating a \texttt{Molecule} object with the properties mentioned above allows us to use JAX to implement differentiable functions of such arrays.
Particularly useful is the function \texttt{jax.grad()}, which allows us to differentiate throughout the code. This is one of the central capabilities of our library. For example, it can be used to directly compute $\nabla \rho(\bm{r})$ using automatic differentiation with a few lines of code. An example describing how to do this is available in the repository.\\

The second main component of the library is the \texttt{Functional} class, which provides the functionality needed to implement exchange-correlation functionals.
Defining a functional as indicated in Eq.~\eqref{eq:param-func} requires the specification of two key methods. The first one is \texttt{energy\_densities}, which receives a \texttt{Molecule} instance and computes the energy densities $\bm{e}[\rho]$.
The second method is \texttt{coefficients}, whose output is the vector of coefficient functions $\bm{c}_\theta[\rho]$. If the method requires properties of the molecule as input, for example the density and its derivatives, they can be specified using the function \texttt{coefficient\_inputs}, whose input is again a \texttt{Molecule} instance.\\

The code block below shows an example of how to define the LDA exchange functional, where we use the energy density specified in Eq.~\eqref{eq:LDA_density}, and for simplicity sum the two spin contributions $\sigma\in\{\uparrow, \downarrow\}$.\\

\begin{minted}[bgcolor=LightGray, linenos]{python}
from jax import numpy as jnp
from grad_dft import Functional

def energy_densities(molecule):
    rho = molecule.density()
    lda_e = -3/2 * (3/(4*jnp.pi))**(1/3) * (rho**(4/3)).sum(axis=1,keepdims=True)
    return lda_e

coefficients = lambda self, rho: jnp.array([[1.]])

LDA = Functional(coefficients, energy_densities)
\end{minted}

In the example above, the \texttt{coefficients} function always returns $1$, which gets broadcasted over the dimensions of the energy density. More generally, as discussed in Sec.~\ref{sec:parametrized_functionals}, it can be substituted with a neural network to implement a neural functional. Our library has a dedicated class \texttt{NeuralFunctional} that provides additional functionality, including the saving and loading of checkpoints, as well as parameter initialization. We present an example of a neural functional below, where inputs to the neural network are the electronic density $\rho_\sigma$ and the kinetic energy density $\tau_\sigma$. We reuse the LDA energy density defined above. Since this functional requires a single energy density, we select the parameter \texttt{features} = 1 in the dense layer.\\

\begin{minted}[bgcolor=LightGray, linenos]{python}
from jax.nn import sigmoid
from flax import linen as nn
from grad_dft import NeuralFunctional

def coefficient_inputs(molecule):
    rho = molecule.density()
    kinetic = molecule.kinetic_density()
    return jnp.concatenate((rho, kinetic))

def coefficients(self, rhoinputs):
    x = nn.Dense(features=1)(rhoinputs)
    x = nn.LayerNorm()(x)
    return sigmoid(x)

neuralfunctional = NeuralFunctional(coefficients, energy_densities, coefficient_inputs)
\end{minted}
The weights and biases of the neural network may then be initialized providing a random key and the inputs.\\
\begin{minted}[bgcolor=LightGray, linenos]{python}
from jax.random import PRNGKey
key = PRNGKey(42)
cinputs = coefficient_inputs(molecule)
params = neuralfunctional.init(key, cinputs)
\end{minted}

JAX may sometimes display numerical instabilities when dealing with numbers of vastly different orders of magnitude. Thus, the user is encouraged to use \texttt{jnp.clip}, \texttt{jnp.round}, and logarithms to circumvent these issues, as we have done throughout the library. Functionals may also be initialized with energy densities or coefficient inputs whose derivatives are not meant to be calculated using automatic differentiation, for instance due to numerical stability or memory issues. Their derivatives with respect to the reduced density matrix should then be hand-defined by the user. In particular, we use the approach proposed in Ref.~\cite{dm21}, to compute the Hartree-Fock exchange energy, whose gradient would otherwise have an unacceptably large memory footprint.\\

The class \texttt{Functional}, and thus \texttt{NeuralFunctional} too, incorporate additional methods. One such example is \texttt{neuralfunctional.energy()}, which can be used to predict the total energy of a molecule.
\begin{minted}[bgcolor=LightGray, linenos]{python}
predicted_energy = neuralfunctional.energy(params, molecule)
\end{minted}
 JAX allows us to automatically differentiate the exchange \& correlation contribution to the energy calculation with respect to the reduced density matrix $\Gamma$, obtaining the Fock matrix, $F^{\sigma}_{\text{KS}}$:
 \begin{equation}
     F_{\text{KS};pq}^{\sigma} = F_{\text{core};pq} + F_{\text{Ha};pq} + \frac{\partial E_{xc,\theta}}{\partial\Gamma_{pq}^\sigma},
 \end{equation}
where $p$ and $q$ index atomic orbitals, $F_{\text{core};pq}$ is a core Fock matrix element (the kinetic and external parts) and $F_{\text{Ha};pq}$ is a Coulomb Fock matrix element. The core and Hartree terms were precomputed as ingredients for calculating the total energy using Gaussian basis sets. The calculation of the Fock matrix is shown in the code below.\\
\begin{minted}[bgcolor=LightGray, linenos]{python}
from jax import grad

def compute_energy(rdm1, molecule):
    molecule = molecule.replace(rdm1=rdm1)
    return functional.energy(params, molecule)

fock_matrix = grad(compute_energy, argnums=0)(molecule.rdm1, molecule)
\end{minted}
For the case of a functional requiring energy densities or inputs to the neural network that cannot be automatically differentiated, we provide the function \texttt{predict\_molecule} that handles the computation of the Fock matrix.\\

In addition, we can take derivatives of the parameters to train the functional. In the following example, we use the JAX \texttt{value\_and\_grad} method as a decorator of the square error loss function \texttt{SEloss} to compute the gradient of the output (\texttt{loss}) with respect to the first argument (\texttt{params}).\\

\begin{minted}[bgcolor=LightGray, linenos]{python}
from tqdm import tqdm
from jax import value_and_grad
from optax import adam, apply_updates
from grad_dft import energy_predictor

# Define optimizer
n_epochs, learning_rate, momentum = 20, 1e-5, 0.9
optimizer = adam(learning_rate=learning_rate, b1=momentum)
opt_state = optimizer.init(params)
compute_energy = energy_predictor(neuralfunctional)

# Define loss
@value_and_grad
def SEloss(params, compute_energy, molecule, trueenergy):
    predictedenergy, fock_matrix = compute_energy(params, molecule)
    return (predictedenergy - trueenergy) ** 2

# Perform training
for iteration in tqdm(range(n_epochs), desc='Training epoch'):
    loss, grads = SEloss(params, compute_energy, molecule, trueenergy)
    updates, opt_state = optimizer.update(grads, opt_state, params)
    params = apply_updates(params, updates)
\end{minted}

Grad DFT also provides a ready-to-use implementation of the self-consistent field (SCF) iterative procedure \texttt{scf\_loop}. \texttt{scf\_loop} is initialized with a functional, and other optional parameters such as the number of iterations or the convergence criteria. The resulting \texttt{scf\_iterator} can then be used with any molecule.
\begin{minted}[bgcolor=LightGray, linenos]{python}
from grad_dft import scf_loop, diff_scf_loop
scf_iterator = scf_loop(neuralfunctional)
updated_molecule = scf_iterator(params, molecule)
print('Energy predicted self-consistently:', updated_molecule.energy)
\end{minted}
The function \texttt{scf\_loop} may be substituted by the end-to-end differentiable and just-in-time compiled alternative \texttt{diff\_scf\_loop}, if the functional does not contain exact exchange terms. These components require the recalculation of two-electron integrals, which remain dependent on PySCF and are therefore not differentiable. Save for this restriction, we can slightly modify the above code to make the training self-consistent. To put it differently, we can train using the predicted energies from the self-consistent iteration.\\
\begin{minted}[bgcolor=LightGray, linenos]{python}
# Define optimizer
n_epochs, learning_rate, momentum = 20, 1e-5, 0.9
optimizer = adam(learning_rate=learning_rate, b1=momentum)
opt_state = optimizer.init(params)
scf_iteration =  diff_scf_loop(neuralfunctional, cycles=30)

# Define loss
@value_and_grad
def SCFloss(params, scf_iteration, molecule, trueenergy):
    updated_molecule = scf_iteration(params, molecule)
    return (updated_molecule.energy - trueenergy) ** 2

# Perform training
for iteration in tqdm(range(n_epochs), desc='Training epoch'):
    loss, grads = SCFloss(params, scf_iteration, molecule, trueenergy)
    updates, opt_state = optimizer.update(grads, opt_state, params)
    params = apply_updates(params, updates)
\end{minted}
These self-consistent iterative procedures are issued with direct inversion in the iterative subspace (DIIS) methods~\cite{pulay1980convergence,pulay1982improved}, that significantly helps converge the self-consistent iteration.\\

We also provide an alternative procedure consisting of direct optimization of the molecular orbitals~\cite{li2023dft}, initialized with \texttt{mol\_orb\_optimizer}. Its corresponding JIT-compilable counterpart, \texttt{jitted\_mol\_orb\_optimizer}, is also available, with similar restrictions. 
In addition, the library allows defining possibly neural non-local correlation functionals using class \texttt{DispersionFunctional} of the form of dispersion tails, as in~\eqref{eq:dispersion}  and~\eqref{eq:damping_dispersion}. They can be passed to \texttt{molecule\_predictor} for joint energy predictions.
We provide an example of this case in the GitHub repository. In addition to the squared energy loss shown above, Grad DFT also provides loss functions to train against the electronic density and a combined loss using both the energy \textit{and} electronic density as proposed in \cite{li2021kohn}.\\

Finally, since constraints of the exact exchange-correlation functional have historically proven useful in the design of accurate approximations~\cite{SCAN,kaplan2023constraints}, we also implement fifteen of them in the form of loss functions. For example, a well-known constraint states that the exchange energy of any electronic density should behave as the average exchange energy of the unpolarized versions of its spin components,
\begin{equation}
    E_x[\rho]  = \frac{1}{2}(E_x[2\rho_\uparrow] + E_x[2\rho_\downarrow]).
\end{equation}
This constraint is named X2 in Ref.~\cite{kaplan2023constraints}. We can compute the corresponding loss with the following code:\\
\begin{minted}[bgcolor=LightGray, linenos]{python}
from grad_dft import constraints
loss = constraints.x2(neuralfunctional, params, molecule)
\end{minted}
These constraints often refer to either the exchange or correlation components of the functional. Consequently, we allow the user to define a \texttt{functional.exchange\_mask} attribute indicating which elements in the $\bm{e}[\rho]$ vector are of type exchange, as opposed to correlation. \\

Overall, our software is versatile and supports differentiable prototyping and experimentation with neural functionals. It includes helpful auxiliary functions such as a differentiable self-consistent iterative procedure that may be used to train or evaluate a neural functional. In the GitHub repository, we also provide examples of different difficulty levels and the code replicating some of the experiments described in the next section. The correctness of the software is monitored by integration tests validating the implementation of several functionals and the self-consistent iteration.\\

Finally, our library enables the use of transfer learning techniques to profit from already trained neural functionals. Loading the parameters into the dictionary \texttt{params} and replicating the original architecture are sufficient to use any pre-trained functional for inference or fine-tuning. In the repository, we provide an example of loading the weights of the DM21 functional and testing its correct implementation~\cite{dm21}.

\subsection{A diatomic molecule energy dataset\label{ssec:dataset}}

\begin{figure}
    \centering
    \includegraphics[width=0.9\textwidth]{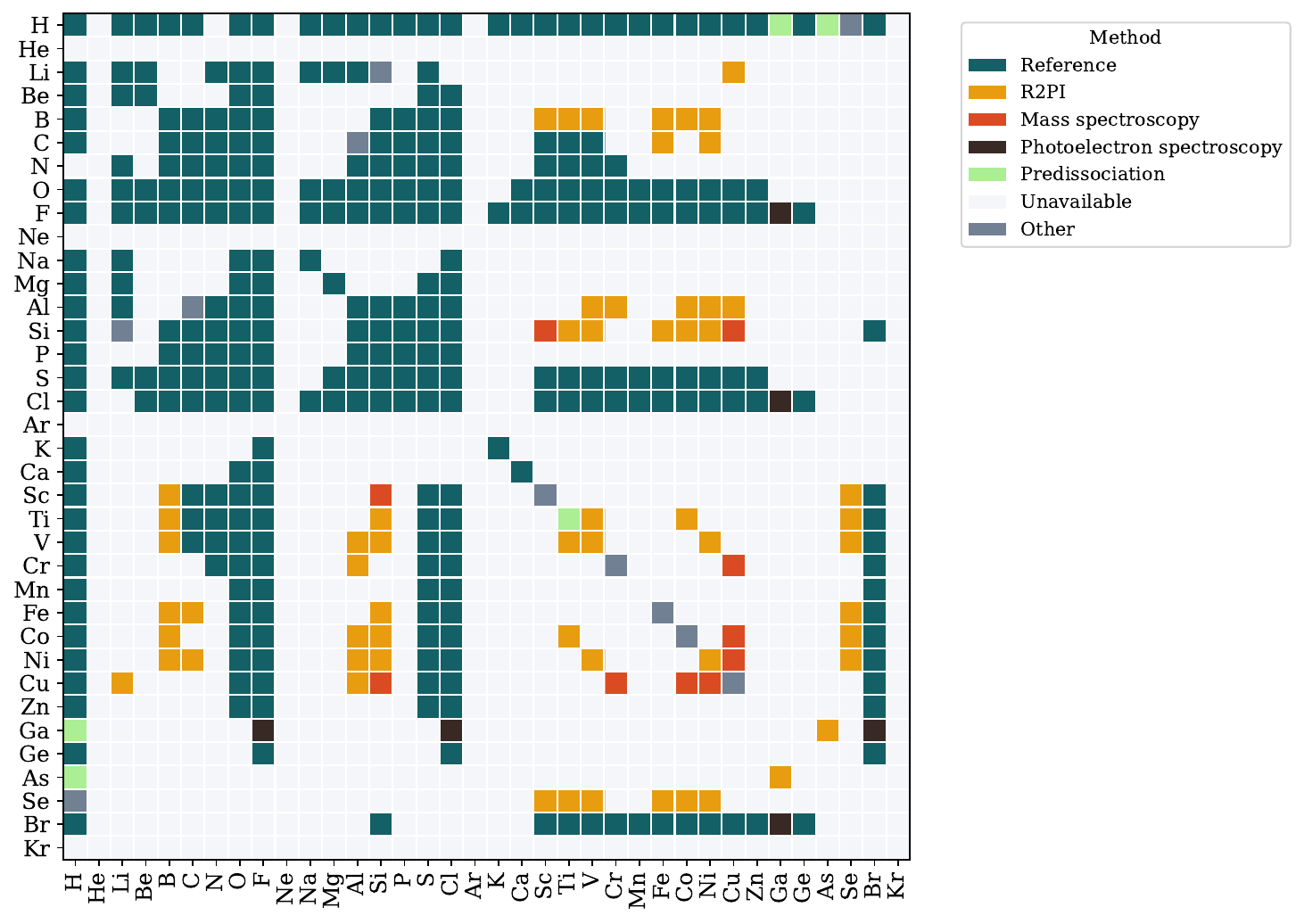}
    \caption{Methods used to compute the dissociation energy of the dimers in the dataset. Colored squares indicate a dimer containing the elements from the corresponding row and column. `Reference' means the data is obtained from reference datasets, such as Ref.~\cite{crchandbook}. `R2PI' stands for resonant two-photon ionization.}
    \label{fig:dataset}
\end{figure}

In addition to the software library, we provide a benchmark dataset of ground state energies of dimers with atoms up to atomic number $Z=36$. The data is obtained from a literature compilation of experimental dissociation energies from Refs.~\cite{hait2019levels,moltved2018chemical,ouyang2008first,jensen2007performance,merriles2019bond,merriles2021chemical,tzeli2008electronic,wu2007electronic,matthew2017determination,sorensen2016bond,sorensen2020bond,wu2006electronic,sevy2017bond,sevy2018bond,harrison2000electronic,gutsev2004periodic,schultz2005databases,ruette2005diatomic,singh2005spectroscopic,li2013diatomic,balasubramanian1989spectroscopic,lemire1990spectroscopy}.
The electronic energy of the diatomic molecules may be decomposed as~\cite{gutsev2004periodic}
\begin{equation}
    E_{\text{molecule}} = D_{0} + E_0 + E_{\text{atom}_1} + E_{\text{atom}_2}.
\end{equation}
For each diatomic molecule, the present dataset contains its experimental dissociation energy $D_0$, the spin, and the experimentally measured (or if not available, DFT calculated) inter-atomic distance and zero point energy $E_0$. The latter is computed from the vibrational frequencies $\omega$ reported in the literature as $E_0 \approx \omega/2$. Since this is also a relatively small correction, we neglect the induced error due to using computational data and the approximate expression for $E_0$. We also neglect relativistic contributions, as they are even smaller and frequently unavailable, except for a few dimers in Ref.~\cite{moltved2018chemical}.
Single-atom energies are also provided. They were computed with CCSD(T)/CBS in PySCF~\cite{sun2018pyscf,sun2020recent} using the two and three-point extrapolation methods from Ref.~\cite{lesiuk2019complete}, with basis sets cc-pVDZ, cc-pVTZ and cc-pVQZ~\cite{dunning1989gaussian,wilson1999gaussian,woon1993gaussian,prascher2011gaussian,balabanov2005systematically}. We also provide similar calculations for cc-pV5Z and cc-pV6Z, although they are not available for all elements.\\

It is worth mentioning that the dissociation energy reported in the literature varies between $D_0$ and $D_e = D_0 + E_0$. 
We provide $D_0$ either as reported from experiments, or derived from $E_0$ and $D_e$ if the latter is experimentally measured.
The present dataset includes 107 dimers with at least one transition metal atom and 111 dimers devoid of them. The experimental methods used to measure the dissociation energy are indicated in Fig.~\ref{fig:dataset}.

\section{Numerical Experiments\label{sec:experiments}}

\begin{figure}
    \centering
\includegraphics[width=\textwidth]{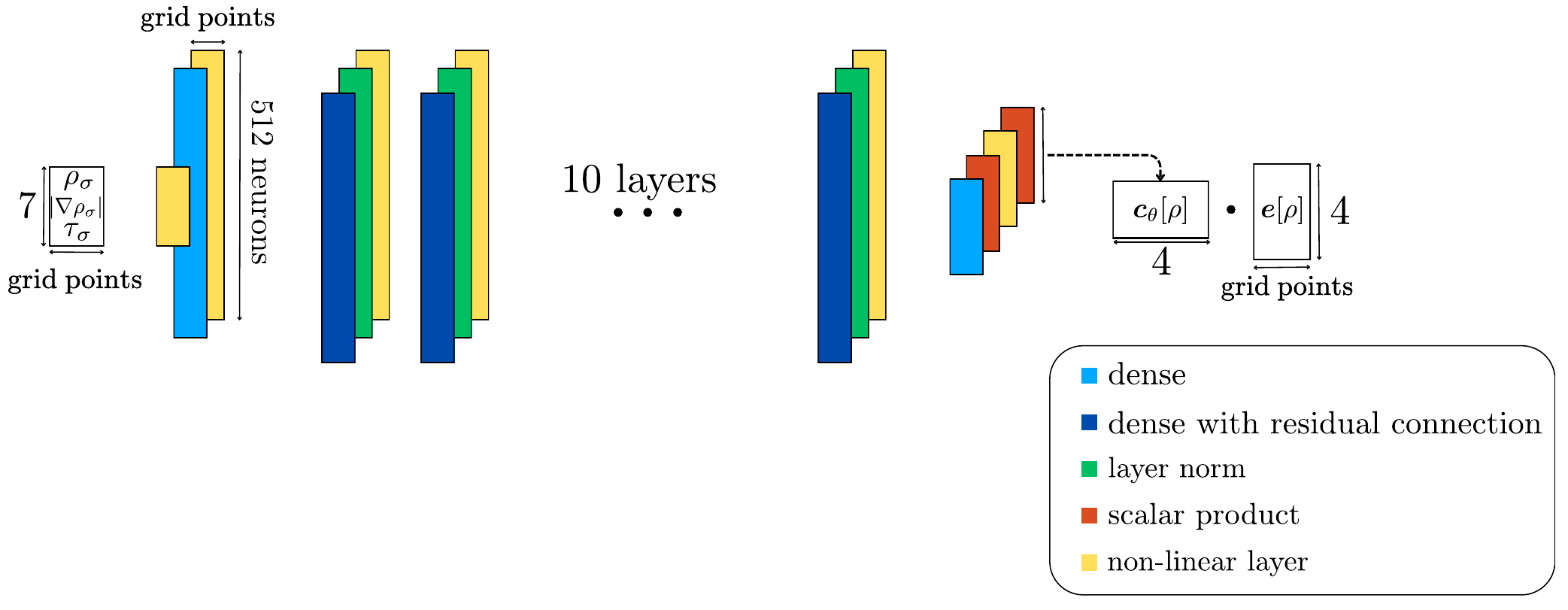}
    \caption{Neural functional used in the experiments, with the inhomogeneous correction factor of~\eqref{eq:ICF} with $N = M = 1$, and the variables defined in~\eqref{eq:experimental_variables}. The neural functional takes as input features of the electronic density, specifically $\rho_\sigma$, $|\nabla \rho_\sigma|^2$, $|\nabla (\rho_{\uparrow}+\rho_{\downarrow})|^2$ and $\tau_\sigma$. From these features, the neural network computes the vector of coefficient functions $\bm{c}_\theta[\rho]$, which is then used to weigh the electronic densities $\bm{e}[\rho]$ of type meta-GGA used to construct the functional.}
    \label{fig:experimentalmodel}
\end{figure}

This section showcases the versatility of the Grad DFT library with a set of preliminary numerical experiments analyzing the capabilities of neural functionals. The first set of experiments, described in Sec.~\ref{ssec:strong_correlation}, studies how well simple neural functionals can predict unseen energies along a potential energy surface. Molecules in our experiments are chosen to be small to reduce computational resources, but they still exhibit the effects of strong electronic correlations, especially at the dissociation limit. In Sec.~\ref{ssec:atoms_generalize} we similarly study their generalization ability across atomic species. We specifically highlight the capability of our simple neural functional to model transition metals, when trained on a dataset containing them. This is a more ambitious endeavor where we train a neural functional on our tailored dataset. The final experiment, reported in Sec.~\ref{ssec:high-quality-data}, investigates the role of noise in training data and its connection to training error. We expect this kind of experiment to be informative about the quality of data required to train state-of-the-art neural functionals.\\

Before explaining the experimental results, we briefly describe the functional architecture we employ. We focus on neural functionals, where the coefficient functions $\bm{c}_{\theta}(\bm{r})$ are obtained from a neural network. Specifically, we use 10-layer fully connected feed-forward neural networks, with 512 or 1024 neurons per layer depending on the experiment, and residual connections, see Fig.~\ref{fig:experimentalmodel}. We select a pure exchange meta-GGA functional based on~\cref{eq:ICF}, with $N = M = 1$ and dimensionless variables $u_\sigma (\bm{r})$ and $w_\sigma(\bm{r})$ defined as
\begin{equation}\label{eq:experimental_variables}
    x_\sigma = \frac{|\nabla \rho_\sigma|}{\rho^{4/3}_\sigma},\qquad u_\sigma = \frac{\beta x_\sigma}{1+\beta x_\sigma},
\qquad\qquad
    t^{-1}_{\sigma} =\frac{5}{3(6\pi^2)^{2/3}} \frac{\tau_{\sigma}}{\rho^{5/3}_{\sigma}},
    \qquad  w_\sigma = \frac{\beta t^{-1}_\sigma}{1+\beta t^{-1}_\sigma}.
\end{equation}
The selection of these variables with $\beta = 2^{-10}$ is based on a search over architectures and hyperparameters to minimize the error of the untrained neural functional on single atom energies. For simplicity, the different spin terms are summed over.\\

It is worth highlighting that this choice of variables demonstrates a powerful regularization effect. It stabilizes the self-consistent iterative procedure and prevents it from diverging. Further, the choice of variables $u_\sigma$ and $w_\sigma$, the functional architecture, and the initialization of dense layers close to the identity matrix, make the untrained functional relatively accurate: feeding the converged electronic density produced by $\omega$B97M-V results in an average error under 2 Ha over the 36 lightest elements of the periodic table (mean average error (MAE) = 1.3 Ha, mean square error (MSE) = 5.6 Ha$^2$). This is an order of magnitude lower than the alternative architectures we tested.\\

However, this choice of variables and architecture does not ensure that the self-consistent iterative procedure will return accurate energies. We unsuccessfully tried to solve this issue by training a few variations of the functional proposed here, including the use of exact exchange components and the regularization loss function proposed in Ref.~\cite{dm21}. While the self-consistent procedure converged for most molecules, it often returned worse energy estimations than those provided by single estimations of the energy from the converged electronic density generated by another functional ($\omega$B97M-V). This is the same functional employed to generate the electronic densities, used in turn to non-self-consistently train the neural functional of our experiments.

In consequence, for the exploratory experiments below, we resort to single evaluations of the energy from converged electronic densities produced by $\omega$B97M-V, without the self-consistent procedure~\cite{wB97M-V}. Future work will address this problem under the lens of self-consistent training, that is, training the model parameters with predictions generated by a self-consistent iterative loop. Such training has been found to have a powerful regularizing effect~\cite{li2021kohn,kalita2022well}, but is also significantly more computationally expensive. We also use the def2-TZVP basis set and PySCF grid level 2 in all our DFT calculations~\cite{def2basisset}, but tailor the training schedule to the specific experiment.

\begin{figure*}
    \centering
    \includegraphics[width=0.48\textwidth]{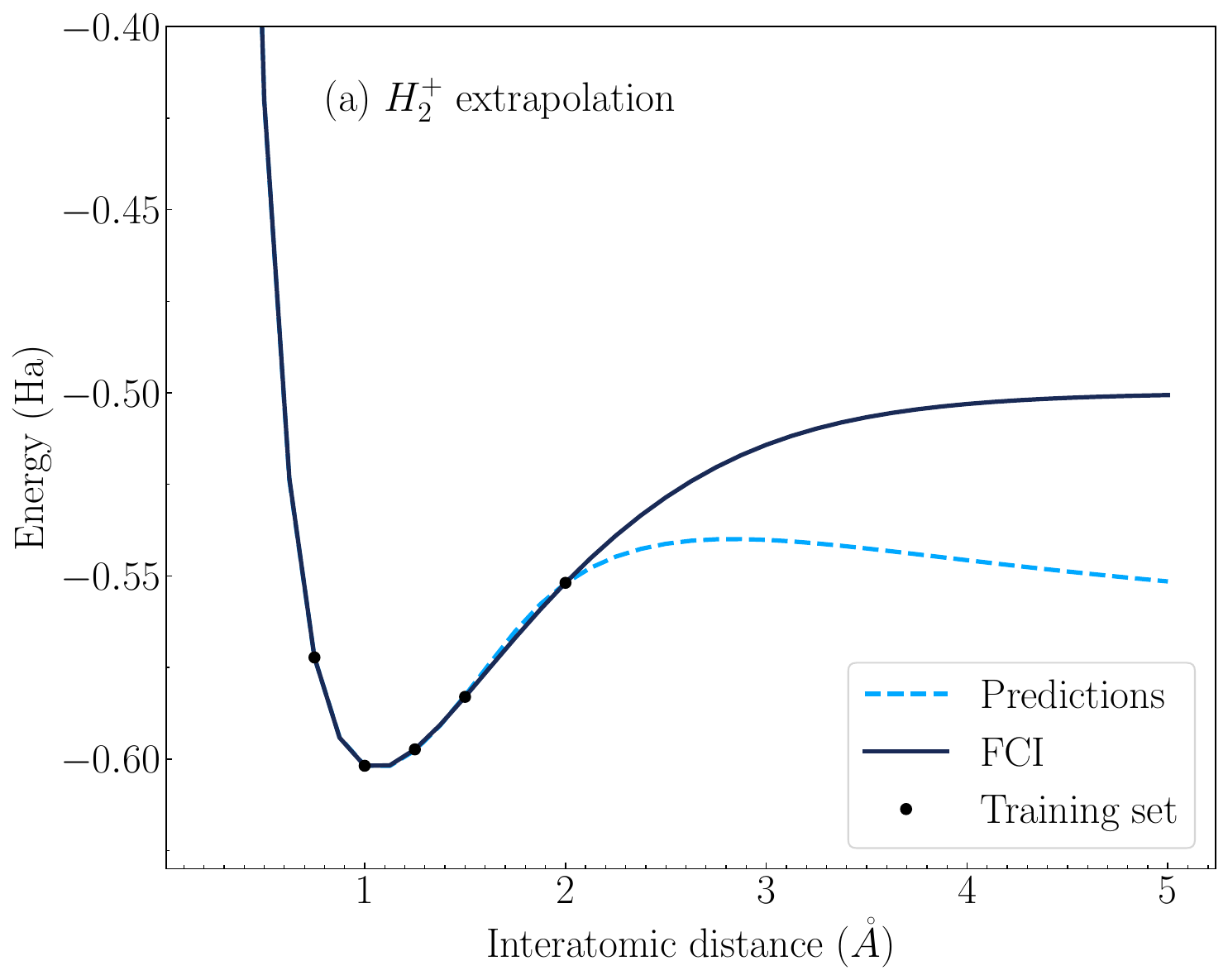}
    \includegraphics[width=0.48\textwidth]{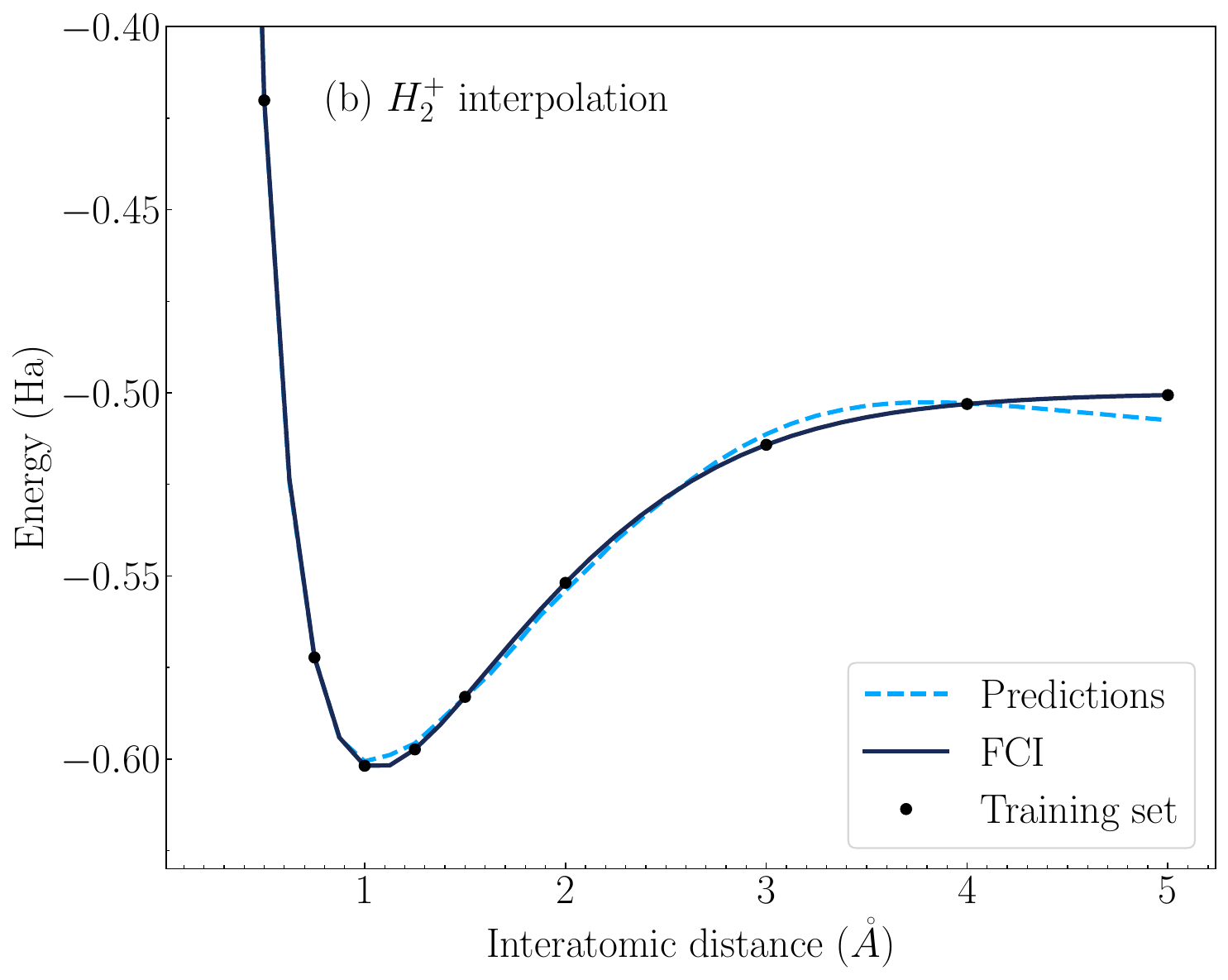}
    \includegraphics[width=0.48\textwidth]{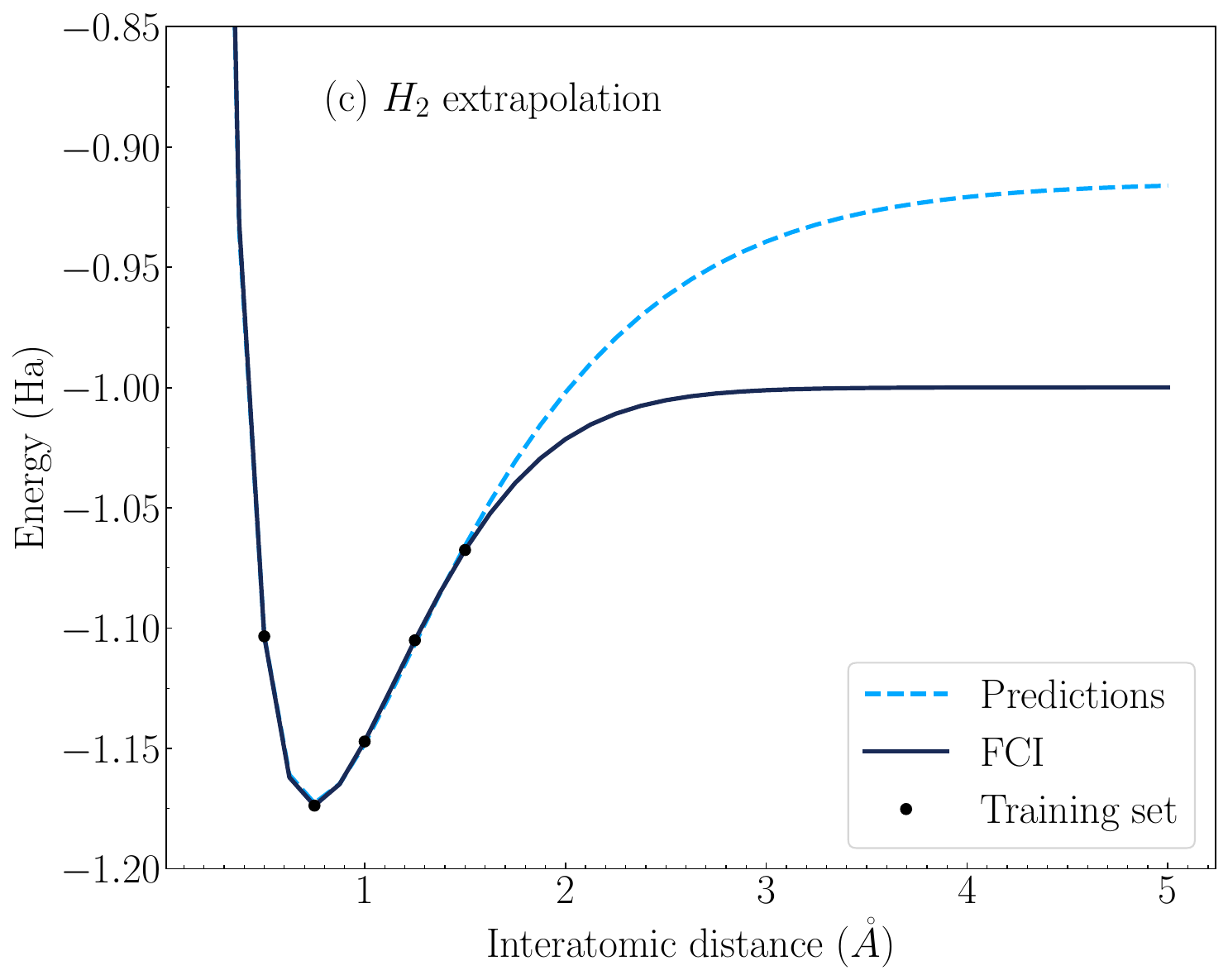}
    \includegraphics[width=0.48\textwidth]{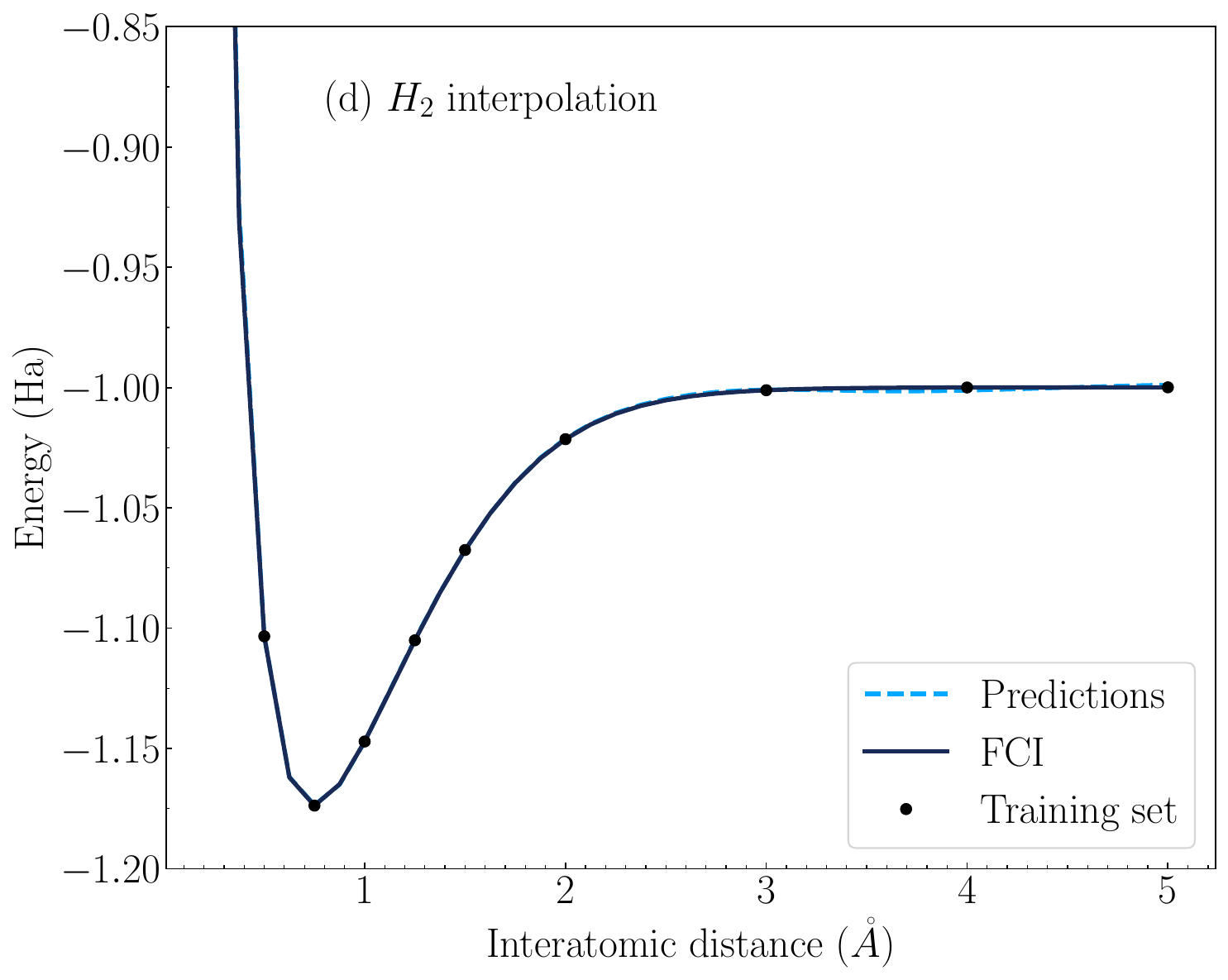}
    \includegraphics[width=0.48\textwidth]{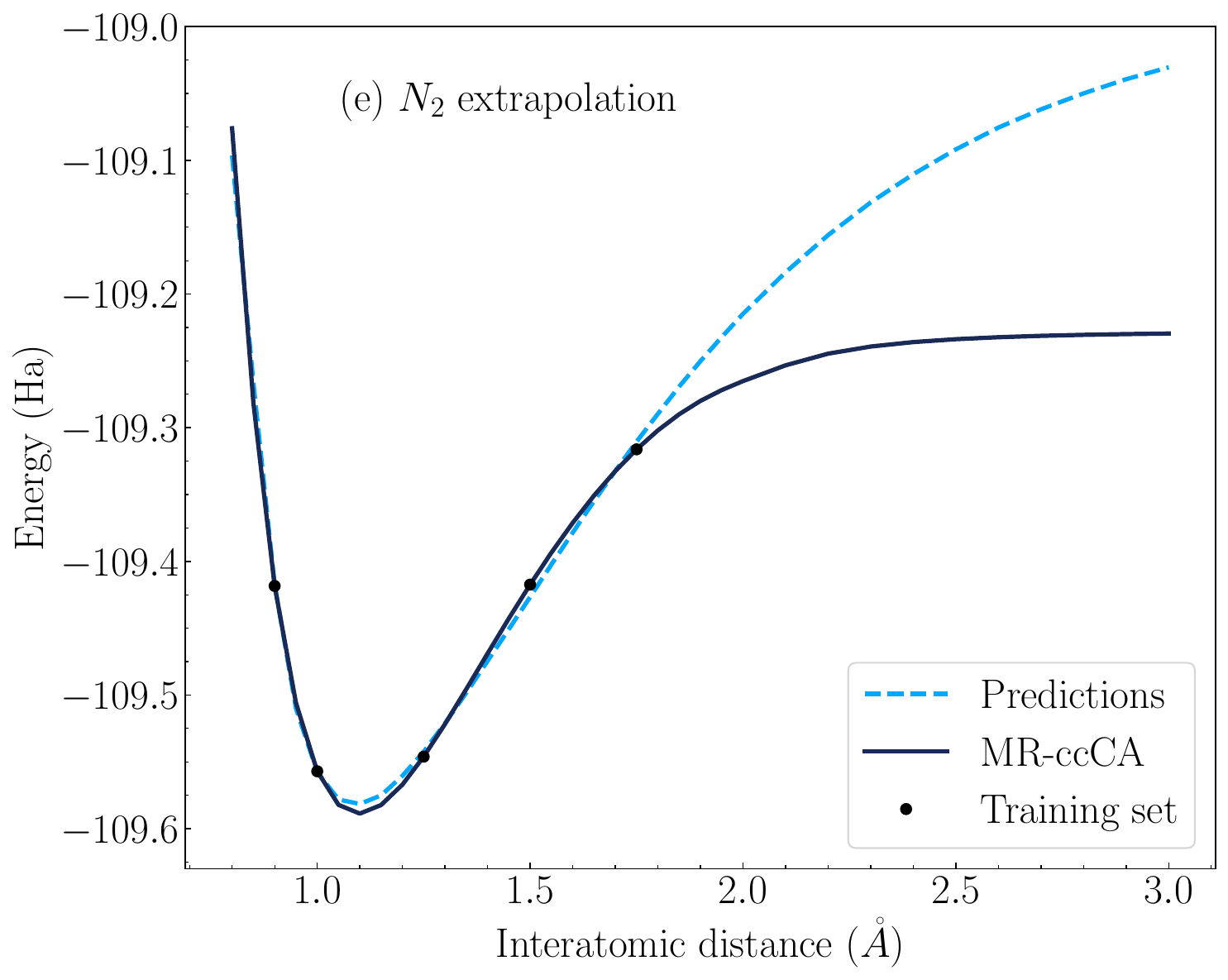}
    \includegraphics[width=0.48\textwidth]{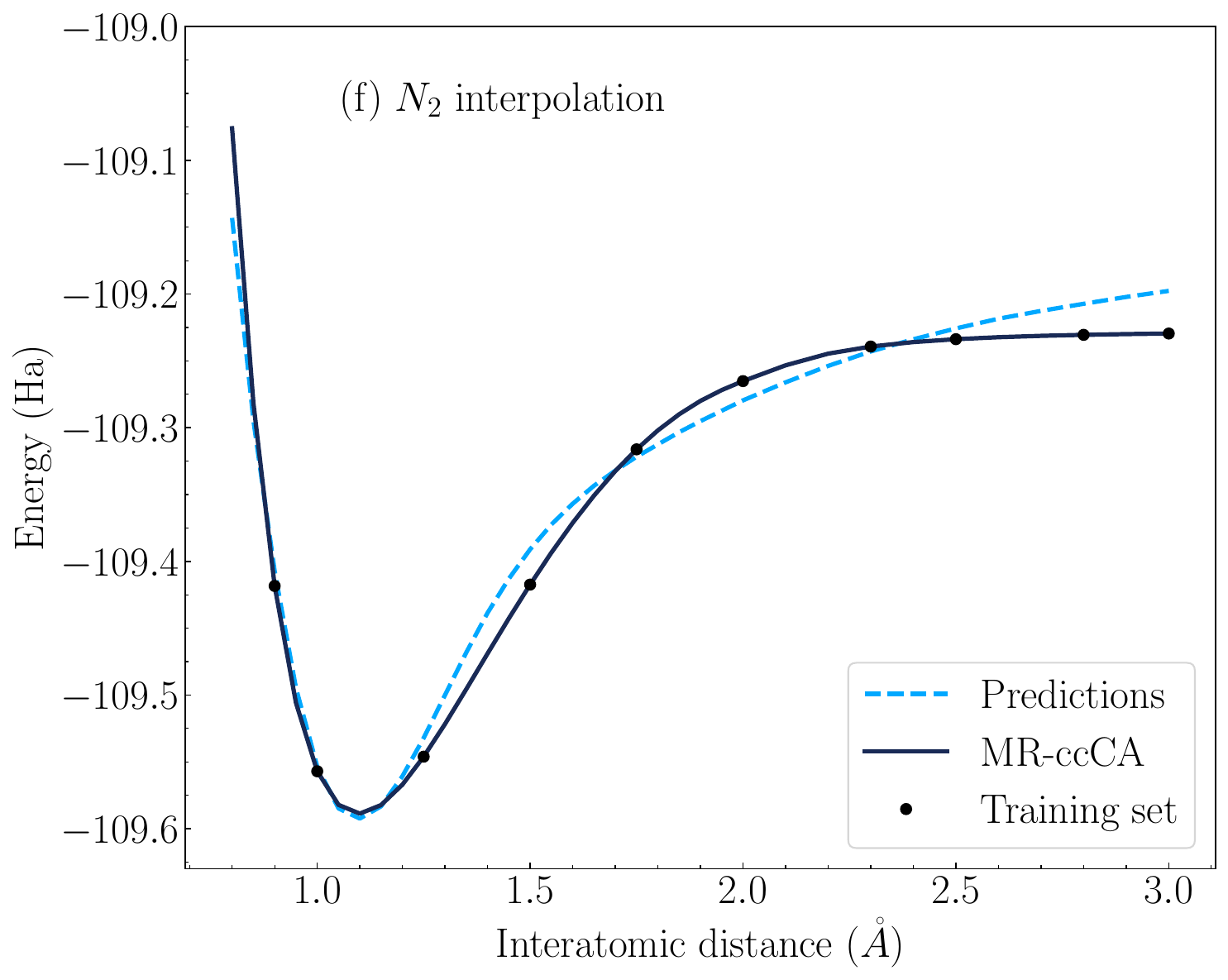}
    \caption{
    Modeling dissociation curves: a neural meta-GGA exchange functional fails to correctly extrapolate the dissociation energies of (a) $H_2^+$ (b) $H_2$ or (c) $N_2$, displaying common systematic DFT biases: over-binding in $H_2^+$ (delocalization error) and under-binding in $H_2$ and $N_2$ (static correlation). The model predictions improve significantly, however, as they are trained on the dissociation regime of (d) $H_2$, while still showing some uncorrected systematic bias in the dissociation of (b) $H_2^+$ and (f) $N_2$. The $N_2$ data was computed in Ref.~\cite{mintz2009computation} using Multi-Reference correlation consistent Composite Approach (MR-ccCA), while 
    we calculated the $H_2$ and $H_2^+$ data using Full Configuration Interaction (FCI) in the cc-pVQZ basis and PySCF.
    }
    \label{fig:dissociation}
\end{figure*}

\subsection{\label{ssec:strong_correlation} Potential energy surfaces}

In this section, we present a few experiments studying the ability of neural functionals to reproduce unseen energies along potential energy surfaces of molecules. Even for very simple systems, stretching molecules reveals some of the historical challenges of density functional theory in dealing with strong correlations. Ref.~\cite{cohen2012challenges} lists two systematic errors of local functionals that can be jointly described as strong correlation: (i) the delocalization error, related also to the self-interaction error, and (ii) the failure to correctly model the static correlations. The former is related to the functional artificially delocalizing the electronic density for a stretched system and the latter stems from standard approximate exchange \& correlation functionals violating the known exact constraints of the exact functional~\cite{cohen2012challenges}.
Among their symptoms, they result in a systematic bias when computing the energy of systems displaying fractional charge and spin effects. In contrast, their correct behavior can be unified under the so-called `flat-plane' condition~\cite{mori2009discontinuous}. A paradigmatic instance of these challenges is the usual failure modes of functionals describing the dissociation of the two simplest molecules: $H_2^+$ and $H_2$.\\

Fig.~\ref{fig:dissociation} shows the performance of the functional modeling the dissociation of $H_2^+$, $H_2$, and $N_2$. We show an extrapolation experiment, where the model is trained on just a few points near the equilibrium geometry; and an interpolation experiment, where it is trained on a more even distribution of points across the entire curve. We observe that our model does not correctly extrapolate the energy in the dissociation regime for any of the molecules, displaying the usual bias, overestimating the binding in the case of $H_2^+$ (delocalization error) and underestimating it for $H_2$ and $N_2$ (static correlation). However, training on stretched geometries significantly improves the resulting predictions, especially in the $H_2$ dimer. Consequently, these experiments suggest that neural functional may help reduce systematic errors in DFT for stretched molecules, if sufficient training data in the dissociated regimes is available. A script to replicate these experiments is included in the examples folder of the GitHub repository.

\subsection{\label{ssec:atoms_generalize} Generalization across atomic species}

\begin{figure*}
\centering
\includegraphics[width=\columnwidth]{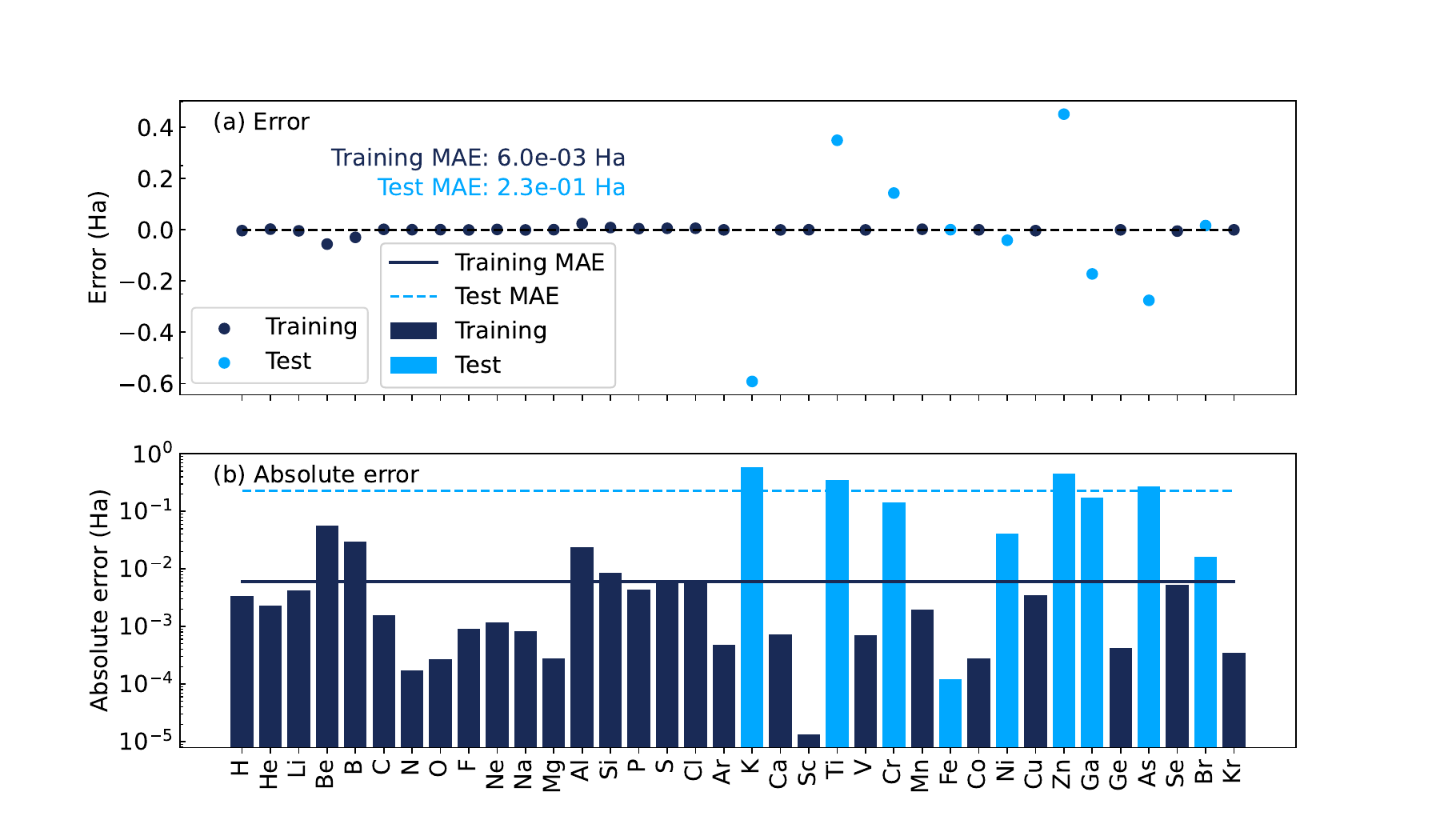}
\caption{Experiment showing that training on a subset of light atoms does not guarantee good prediction accuracy outside the training set. The top plot (a) depicts the signed error of the model predictions, while the bottom one (b) focuses on the absolute error. The mean absolute error (MAE) is significantly higher in the test dataset (light blue bars), by more than an order of magnitude. 
These results highlight the importance of training neural functionals in a variety of systems.
In contrast, transition metals predictions do not appear to be particularly inaccurate. This observation, together with the experiment indicated in~\cref{fig:dimer_tm_heatmap}, highlights the potential of neural functionals to model otherwise challenging electronic systems.
}
\label{fig:atoms_generalization}
\end{figure*}

\begin{figure*}
\includegraphics[width=0.435\columnwidth]{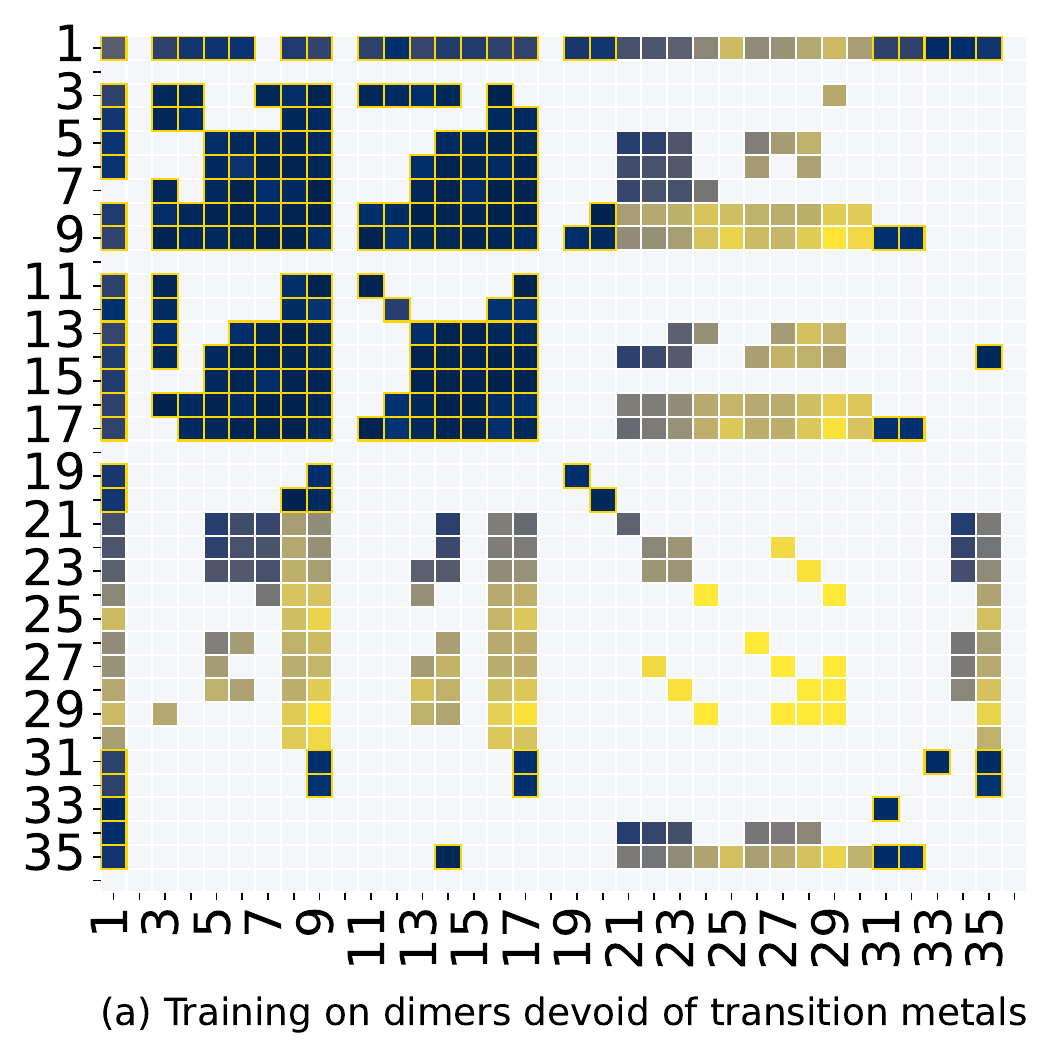}\includegraphics[width=0.56\columnwidth]{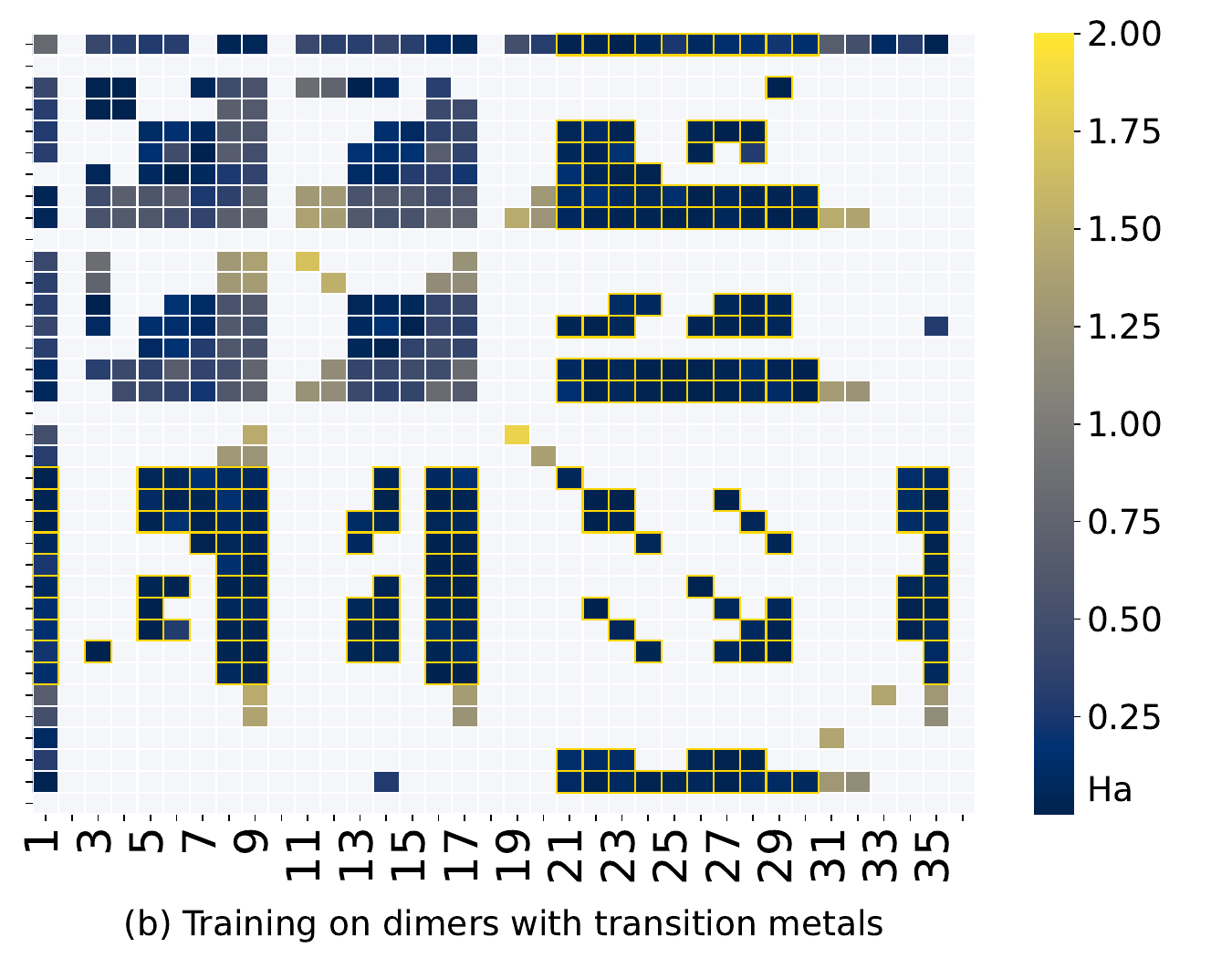}\hspace{10pt}
    \caption{Example of the accuracy (error in Hartrees) of our meta-GGA functional when trained over either (a) non-transition-metal dimers or (b) transition-metal dimers. Each square represents a diatomic molecule composed of atoms whose atomic numbers are indicated in the x and y axes. The molecules over which the model was trained are indicated by a yellow border around the square representing the diatomic molecule. This experiment reinforces the conclusions from the previous one: generalization to unseen molecules remains a challenge, particularly when we aim to generalize to molecules containing transition metals. In this situation, subfigure (a), the model has a training MAE of $0.117$ Ha vs a test MAE of $1.355$ Ha. Surprisingly, if we instead train on dimers containing transition metals, subfigure (b), the training MAE is even better, $0.058$ Ha, and the test MAE also improves to $0.553$ Ha. As in~\cref{fig:atoms_generalization}, we observe one order of magnitude difference between the training and test MAE in both cases (a) and (b). The training curves of these experiments with 5 random seeds each might be found in~\cref{fig:training} (a).
    }
    \label{fig:dimer_tm_heatmap}
\end{figure*}

Here we study the generalization capabilities of our neural functional across atomic species. First, we analyze its ability to generalize predictions for energies of single atoms up to atomic number $Z=36$.
The functional is built from a feed-forward neural network of 10 layers and 1024 neurons per layer. 
The resulting model is tested on elements in the fourth row of the periodic table: transition metals with even atomic number (Ti, Cr, Fe, Ni, and Zn) and non-transition metals with odd atomic number (K, Ga, As, Br). The remaining lighter-than-Rb atomic species form the training set. The results are shown in Fig.~\ref{fig:atoms_generalization}. \\

A key observation from our experiment is that the mean absolute error on the training set is relatively low, $6\cdot 10^{-3}$ Ha, even for our relatively simple and non-self-consistently trained model. We also note that transition metals do not necessarily exhibit the most challenging behavior. For example, our trained model encounters greater difficulty in accurately predicting the ground state energy of beryllium than that of scandium. However, extending this predictive capability to previously unseen atoms remains a formidable challenge, even when the model has effectively learned from a diverse range of neighboring elements. The order of magnitude of difference between the accuracy of the model on training vs test elements further underscores the importance of training on a diverse set of data.
\\

 This analysis is complemented by a second experiment, using a similar neural functional with 10 layers of 512 neurons each. We consider our tailored dataset introduced in subsection~\ref{ssec:dataset}, and compare two distinct subgroups: the first comprising dimers containing at least one transition metal, and the second consisting of diatomic molecules devoid of such elements. This approach allows us to investigate how the inclusion or exclusion of transition-metal data influences the generalization capabilities of our model.\\

\begin{figure*}
    \includegraphics[width=0.85\columnwidth]{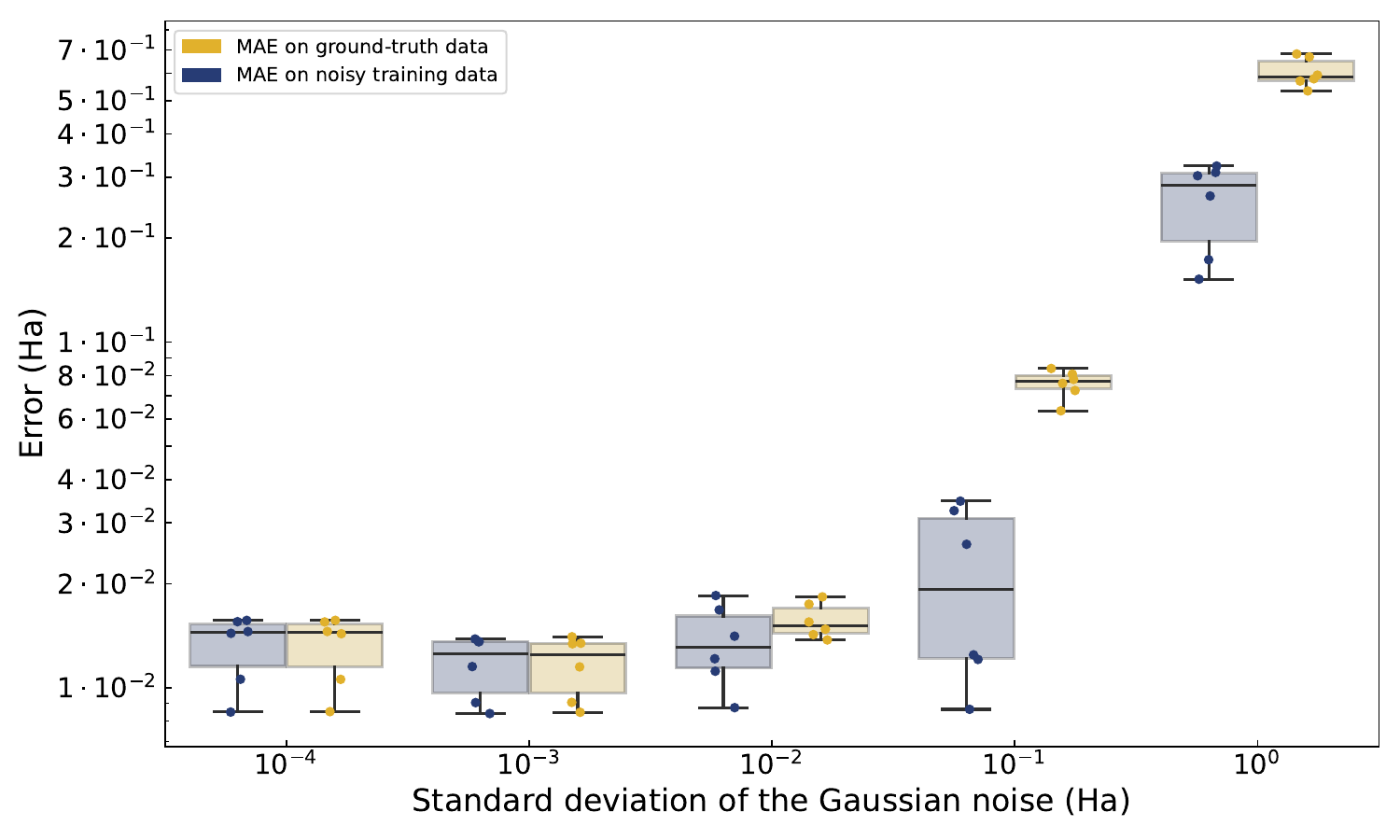}
    \caption{Mean absolute error (MAE) of the electronic energy of single atoms up to atomic number $Z=36$ estimated with our meta-GGA neural functional, with respect to the noisy training data (blue) and the ground truth (gold). Each point represents one randomly initialized and then trained model; with newly independently sampled random noise on the training data.
    The noise distribution introduced is Gaussian centered on 0, with standard deviations $10^{-n}$ Ha, for $n\in[-4, 0]$. 
    We observe that noise equal to or greater than $0.1$ Ha makes the MAE with respect to the training data diverge from the MAE with respect to the true energy. Below that noise level, we observe little gain in the training error from increasing the accuracy of the data.}
    \label{fig:noise_error}
\end{figure*}

\begin{figure}[ht!]
    \centering
    \includegraphics[width=0.85\columnwidth]{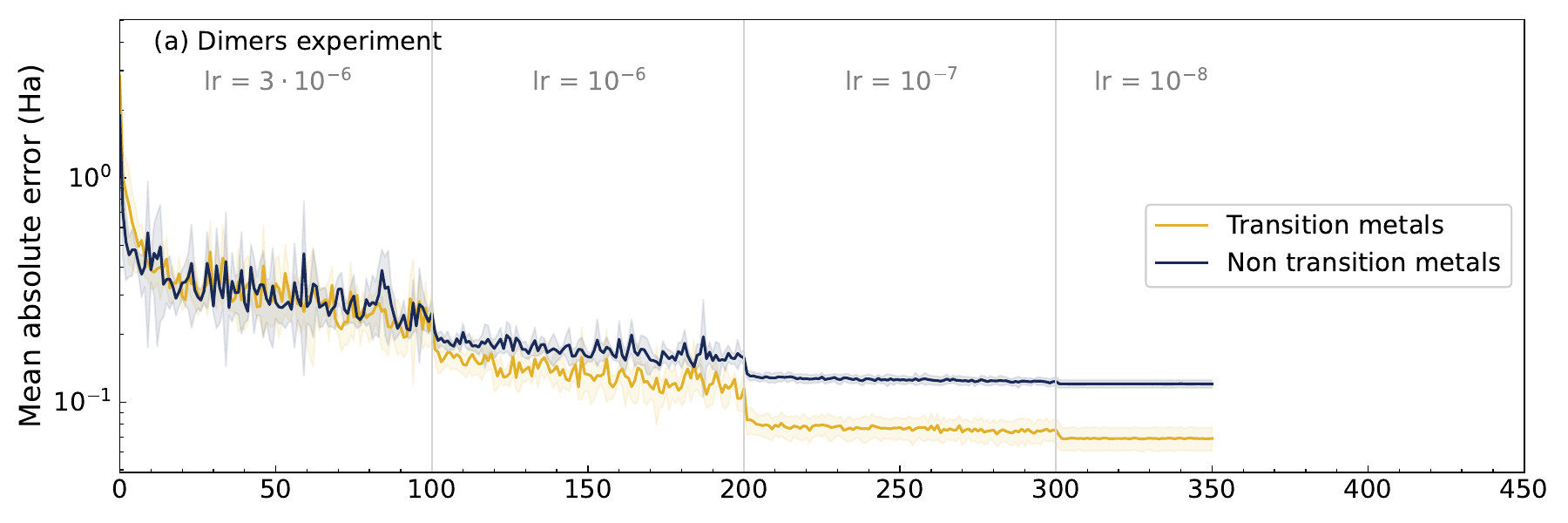}
    \includegraphics[width=0.85\columnwidth]{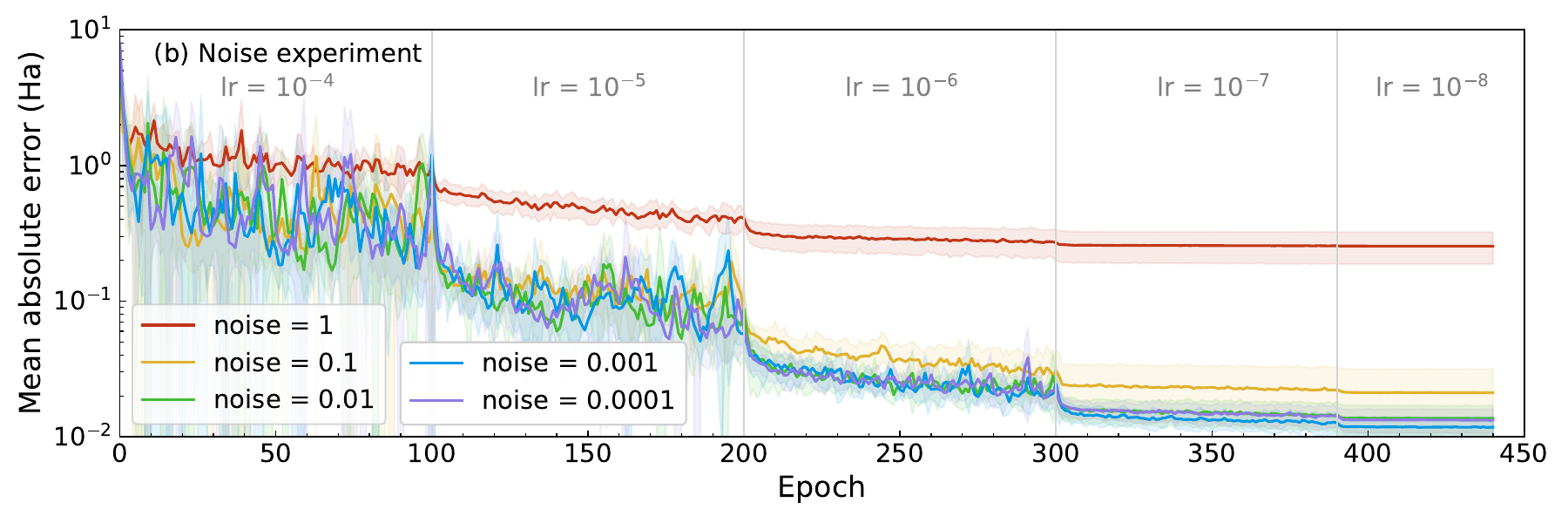}
    \caption{Training loss vs epoch number in (a) the dimer dataset experiments from~\cref{fig:dimer_tm_heatmap} and (b) the noise experiment from~\cref{fig:noise_error}. The training rounds were carried out with a piece-wise learning rate indicated by `lr' in the graph, and the Adam optimizer~\cite{Kingma15Adam}. The architecture is similar in both experiments, the meta-GGA neural functional defined in~\eqref{eq:experimental_variables} with a feedforward neural network of 10 layers. However, they differ in the number of neurons per layer: experiment (a) is more memory intensive so we use 512 neurons per layer, while experiment (b) employs 1024. We use a batch size of 1, since the memory requirements for even a single electronic system are quite high due to the size of the grid. We depict the average and standard deviations computed with 5 random seeds in subfigure (a) and 6 random seeds for each noise level in subfigure (b).}
    \label{fig:training}
\end{figure}

The resulting error is displayed in Fig.~\ref{fig:dimer_tm_heatmap}, and the corresponding training curve in Fig.~\ref{fig:training} (a). As in the previous experiment, shown in~\cref{fig:atoms_generalization}, we highlight the challenge of generalizing to unseen molecules. This difficulty is particularly stark when the training dataset lacks molecules with transition metals but we aim to predict compounds containing them. In such a case, the MAE on the test dataset exceeds 1 Ha. In contrast, transition metals do not appear to be more challenging to learn to model than other species. Specifically, the training MAE on~\cref{fig:dimer_tm_heatmap} (b) is $0.058$ Ha, lower than the training MAE in~\cref{fig:dimer_tm_heatmap} (a), $0.117$ Ha. This result hints at the potential of neural functionals to model complex electronic configurations.

\subsection{Accuracy of training data\label{ssec:high-quality-data}}

One of the motivations for constructing functionals using machine learning techniques is that the trained functionals may be able to achieve unprecedented levels of accuracy. For this reason, training data should ideally be of excellent quality. In this subsection, we perform preliminary experiments testing the impact of training noise on the predictive capabilities of neural functionals. 
More concretely, we study how the training and test errors of the functional change as we introduce Gaussian noise in the training data. We use the meta-GGA neural functional described at the beginning of this section, with 10 layers of 1024 neurons each. As in previous cases, the objective is to learn the energy of isolated atoms lighter than Rb.\\

Since we want to measure the impact of training noise, we test the capability of our model to replicate the behavior of the well-known functional $\omega$B97M-V under noisy training conditions~\cite{wB97M-V}. This implies that, in contrast to the rest of the experiments, the training energies will be computed with the $\omega$B97M-V functional. We will similarly use the same $\omega$B97M-V-converged electronic densities, basis set (def2-TZVP), and grid (PySCF grid level 2). This experimental setup allows us to isolate the impact of noise in the target energy, as the electronic density and other variables are fixed and do not introduce additional sources of errors.\\

The results are indicated in Fig.~\ref{fig:noise_error} and Fig.~\ref{fig:training} (b). They show that training on increasingly accurate data is only valuable up to the point where data noise becomes similar in magnitude to the training error. For example, in fig.~\ref{fig:noise_error}, the training error oscillates around $10^{-2}$ Ha in the low-noise regime.
Below that threshold, the training error remains approximately constant, and the mean absolute errors with respect to the true and noisy training data are close. This suggests that in such a regime increasing the accuracy of the training data is no longer useful: the main contribution to the final error is due to the functional modeling capabilities, not the data. Although preliminary, this experiment reveals an interplay between these two factors. Better data appears to be important only if the model is powerful enough to make good use of it.  \\

\section{Discussion}


For all its remarkable success, DFT still faces significant challenges in modeling strongly correlated quantum systems. Conversely, deep learning is a newer field, but it has already yielded noteworthy breakthroughs in several scientific research areas. This suggests the possibility that it could be similarly applied to the design of highly accurate yet inexpensive neural functionals. Our work introduces Grad DFT, which provides a key tool to assist researchers in advancing the field of machine learning-enhanced density functional theory by rapidly developing, training, and evaluating new functionals.\\

While our library was built to be both versatile and powerful, there are opportunities for enhancements that could further increase its usefulness. For example, it would be highly desirable to enhance the computational performance by improving the interface with multiple GPUs. This would reduce the amount of time required to train neural functionals. 
Moreover, the support for non-local correlation and exact exchange functionals could be upgraded. Specifically, these elements in Grad DFT rely on PySCF and Libcint for the computation of atomic integrals~\cite{sun2015libcint,sun2018pyscf,sun2020recent}. This imposes restrictions on the interoperability between the calculation of the exact exchange energy density and the automatic differentiability of the JAX code used throughout Grad DFT. Finally, the current version of the code does not yet support periodic boundary conditions.\\

Future research should also address the limitations of our experiments. Particularly compelling is the question of whether self-consistent training, which Grad DFT makes possible, can strengthen the regularization of the self-consistent iterative procedure, and help it converge to correct predictions. Additionally, this procedure removes the training data dependence on electronic densities generated by other functionals, and potentially paves the way for a new generation of accurate and inexpensive neural functionals.\\

Notwithstanding future improvements, Grad DFT fills a central role as the software layer of the machine learning-enhanced DFT pipeline we envisaged in the introduction. As discussed throughout this article, Grad DFT displays significant functionality, including a fully differentiable self-consistent iterative procedure. Such capabilities allow researchers to focus on answering the key open questions of the field rather than on software or machine-learning engineering. These questions encompass, from a conceptual perspective, understanding the importance of quality data in the construction of state-of-the-art neural functionals. As such, more work should be devoted to assessing the role of wavefunction methods in generating these data, complementing the experimental results. Quantum computing should also be examined as a technique to generate high-quality data for machine learning-enhanced DFT, which could become an important application for this emergent technology. As hinted in the experiments, these data could have special relevance in ensuring the generalizability of the resulting functionals in otherwise challenging systems. Overall, neural functionals represent a promising approach to greatly enhance the accuracy and applicability of density functional theory. We believe that Grad DFT will significantly simplify research on this exciting new area, and potentially result in valuable insights that help advance this nascent field.

\section{Acknowledgements}
We thank helpful comments and insights from Alain Delgado, Modjtaba Shokrian Zini, Stepan Fomichev, Soran Jahangiri, Diego Guala, Jay Soni, Utkarsh Azad, Kasra Hejazi, Vincent Michaud-Rioux, Maria Schuld and Nathan Wiebe.

This research used resources of the National Energy Research
Scientific Computing Center, a DOE Office of Science User Facility
supported by the Office of Science of the U.S. Department of Energy
under Contract No. DE-AC02-05CH11231 using NERSC award
NERSC DDR-ERCAP0025705.


\section{Code and data availability}
The code and dataset introduced in this paper are openly available in \url{https://github.com/XanaduAI/GradDFT}. The experimental training data and checkpoints are available at \url{https://doi.org/10.5281/zenodo.10280806}.

\bibliography{references}

\end{document}